\documentclass[prx,aps,twocolumn,superscriptaddress]{revtex4}

\usepackage{amsmath}
\usepackage{amsfonts}
\usepackage{amssymb}
\usepackage{graphicx}
\usepackage{color}
\usepackage{chemmacros}
\usepackage{marvosym}
\usepackage{natbib}

\renewcommand{\a}{\alpha}
\renewcommand{\b}{\beta}
\renewcommand{\d}{\delta}
\newcommand{\e}{\epsilon}
\newcommand{\g}{\gamma}
\renewcommand{\k}{\kappa}
\renewcommand{\l}{\lambda}
\newcommand{\s}{\sigma}

\newcommand{\E}{\mathbb{E}}

\newcommand{\N}{\mathcal{N}}
\renewcommand{\L}{\Lambda}

\newcommand{\vs}{\varsigma}
\newcommand{\col}{\color{black}}

\newcommand{\R}{\mathbb{R}}
\newcommand{\beq}{\begin{equation}}
\newcommand{\eeq}{\end{equation}}
\newcommand{\bea}{\begin{eqnarray}}
\newcommand{\eea}{\end{eqnarray}}
\newcommand{\bal}{\begin{align}}
\newcommand{\eal}{\end{align}}

\newcommand{\sm}{\text{\Male}}
\renewcommand{\sf}{\text{\Female}}
\newcommand{\sh}{\text{\Hermaphrodite}}
\newcommand{\sn}{\text{\Neutral}}
\newcommand{\fm}{\text{\FemaleMale}}
\renewcommand{\ss}{\bullet}

\DeclareFontFamily{U}{mathb}{}
\DeclareFontShape{U}{mathb}{m}{n}{
  <-5.5> mathb5
  <5.5-6.5> mathb6
  <6.5-7.5> mathb7
  <7.5-8.5> mathb8
  <8.5-9.5> mathb9
  <9.5-11.5> mathb10
  <11.5-> mathbb12
}{}

\begin{document}

\title{Sex as information processing: optimality and evolution}

\author{Anton S. Zadorin} 
\affiliation{Max  Planck  Institute  for  Mathematics  in  the  Sciences,  Leipzig,  Germany}
\affiliation{Center for Interdisciplinary Research in Biology (CIRB), Coll\`ege de France, CNRS, INSERM, PSL Research University, Paris, France}
\author{Olivier Rivoire} 
\affiliation{Center for Interdisciplinary Research in Biology (CIRB), Coll\`ege de France, CNRS, INSERM, PSL Research University, Paris, France}

\begin{abstract}
{\col The long-term growth rate of populations in varying environments quantifies the evolutionary value of processing the information that biological individuals inherit from their ancestors and acquire from their environment. 
Previous models were limited to asexual reproduction with inherited information coming from a single parent with no recombination. We present a general extension to sexual reproduction and an analytical solution for a particular but important case, the infinitesimal model of quantitative genetics which assumes traits to be normally distributed. We study with this model the conditions under which sexual reproduction is advantageous and can evolve in the context of autocorrelated or directionally varying environments, mutational biases, spatial heterogeneities and phenotypic plasticity. Our results generalize and unify previous analyses. We also examine the proposal made by Geodakyan that the presence of two phenotypically distinct sexes permits an optimal adaptation to varying environments. We verify that conditions exists where sexual dimorphism is adaptive but find that its evolutionary value does not generally compensate for the two-fold cost of males.}
\end{abstract}

\maketitle

\section{Introduction}

Evolution by natural selection relies on the presence of variations which are generated and transmitted through mechanisms that are themselves subject to natural selection. This raises the question of the optimality of these mechanisms in relation to the constraints to which populations are subject. {\col This long-standing problem of population genetics~\cite{kimura1960optimum,levins1967theory,haccou1995optimal,sasaki1995evolutionarily,feldman1996population} can also be approached from the perspective of information theory~\cite{shannon1948mathematical} by viewing the mechanisms for generating and transmitting biological variations as information processing schemes~\cite{Rivoire:2011fy}. The simplest case, where the only source of information is an environmental cue that is sensed and processed to adapt an internal state, corresponds to a model proposed by Kelly in the 1950s to demonstrate how Shannon's theory could be generalized to quantify the value of information~\cite{kelly2011new}. His approach has been applied and extended to quantify the value of biological information for populations of reproducing individuals evolving in varying environments~\cite{bergstrom2004shannon,kussell2005phenotypic,Rivoire:2011fy,rivoire2016informations}. 

When accounting for inherited information, for differences between individuals or for general forms of genotype-to-phenotype maps, the value of biological information cannot be reduced to the entropies originally introduced by Shannon. Instead, differences in long-term growth rates obtained by comparing populations that adopt different information processing schemes provide an appropriate generalization~\cite{Rivoire:2011fy}. This point of view clarifies the diverse modes of adaptation and inheritance that biological organisms exhibit~\cite{lachmann1996inheritance,Rivoire:2014kt,uller2015incomplete,mcnamara2016detection,mayer2016diversity} and leads to multiple analogies with problems and concepts from non-equilibrium statistical physics~\cite{hirono2015jarzynski,kobayashi2015fluctuation,vinkler2016analogy,rivoire2016informations,genthon2020fluctuation}. }


{\col With very few exceptions~\cite{salahshour2019phase,miyahara2019many}, this evolutionary perspective on information processing has been limited to models of vertical inheritance, excluding any form of interaction between members of a same generation. In particular, sex, or genetic exchange, which is a major mode of information transmission in the living world~\cite{bell1982masterpiece} has been left aside. The goal of this article is to show how measures of biological information developed for asexual populations can be extended to sexual populations. We present a general formalism and apply it to solve analytically a central model of quantitative genetics, the infinitesimal model~\cite{lynch1998genetics}. We obtain two results. First, we quantify the value of sexual reproduction over asexual reproduction and thus revisit the long-standing question of the conditions under which sexual reproduction may evolve and be maintained. Second, we quantify the value of sexual dimorphism and thus analyze quantitatively a proposal made by Geodakyan according to which the presence of two sexes permits an optimal adaptation to varying environments~\cite{Geodakyan:1965,Geodakyan:2015}.}



{\col The first question that we analyze,} the conditions under which sexual reproduction can emerge and be maintained, has been extensively studied, although no definite solution is consensually accepted~\cite{smith1978evolution,bell1982masterpiece,otto2009evolutionary,hartfield2012current}. In particular, the origin of the most significant constraints is highly debated~\cite{kondrashov1993classification}. One class of models takes environmental constraints to be determining. It includes notably the red-queen hypothesis~\cite{hamilton1990sexual}, which invokes rapidly varying selective pressures, and the tangled-bank hypothesis~\cite{ghiselin1976economy}, which invokes spatially heterogeneous resources. Another class of models takes genetic constraints to be determining. It includes notably Muller's ratchet~\cite{muller1964relation} and Kondrachov's hatchet models~\cite{kondrashov1988deleterious}. Our formalism integrates both kinds of constraints and provides a new perspective on previously known results~\cite{Charlesworth:1993fg}. In particular, by focusing on the nature of the constraints rather than on their mechanism or origin, it reconciles some of the alternative scenarios. 

The second question {\col that we analyze pertains to the two-fold cost of males in dioecious populations, when assuming that males and females are present at the same ratio. Na\"ively, if males were as fecund as females, the population could double its number of offsprings per generation, which, in terms of growth rate, corresponds to an additional factor $\ln 2$. This suggests that monoecious (hermaphroditic) populations should have a systematic evolutionary advantage over dioecious populations. To explain that many species have nevertheless two distinct sexes, additional constraints are usually integrated, including sexual selection, intersexual food competition or reproductive role division~\cite{hedrick1989evolution}. Alternatively, Geodakyan proposed that two distinct sexes permits an optimal processing of inherited information~\cite{Geodakyan:1965,Geodakyan:2015}. His proposal rests on the assumption that females are developmentally more plastic than males: when the environment changes, their fertility is unaffected, irrespectively of their genotype, while males survive only if they have the most adapted genotypes. This strong selection on males is proposed to favor the integration of the environmental change into the genotypes of the next generation while the weak selection on females mitigates the selection load. Whether the value of this information processing scheme can exceed $\ln 2$ and thus compensate for the two-fold cost of males has, to our knowledge, never been investigated. Our formalism allows us to examine not only the adaptive value of this scenario but also its potential to evolve.} 


\section{Model}

\subsection{Modes of reproduction}

\begin{figure}[t]
\begin{center}
\includegraphics[width=\linewidth]{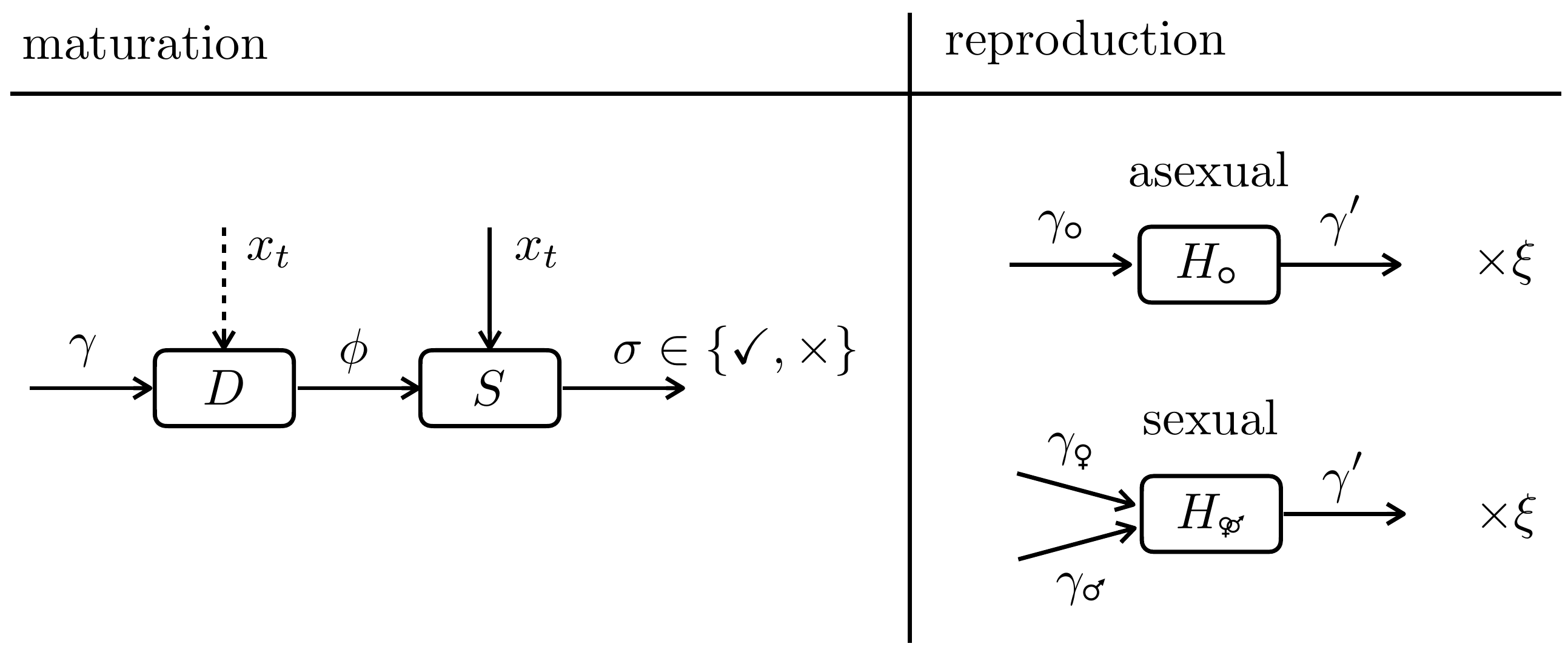}
\caption{Life cycle of individuals in a representation borrowed from information theory, where each ``box'' represents a communication channel, i.e., a conditional probability to generate the output given the input(s)~\cite{Rivoire:2011fy}. First is a maturation step, where an individual with genotype $\g$ acquires a phenotype $\phi$ through development and is selected by the environment $x_t$ based on this phenotype to either survive ($\s=\checkmark$) or die ($\s=\times$). Second, surviving individuals can reproduce. In asexual reproduction, an individual with genotype $\g_\sn$ produces $\xi$ offsprings with genotype $\g'$. In sexual reproduction, a pair of individuals with genotypes $\g_\sf,\g_\sm$ mate to generate $\xi$ offsprings with genotype $\g'$. 
\label{fig:scheme}}
\end{center} 
\end{figure}

We first reformulate a model of asexual reproduction~\cite{Rivoire:2011fy,Rivoire:2014kt} before generalizing it to account for sexual reproduction in monoecious (single sex) and dioecious (two sexes) populations.

\subsubsection{Asexual reproduction}

We assume that at each discrete generation $t$, a newly born individual with genotype $\g$ follows a life cycle consisting of two steps  (Fig.~\ref{fig:scheme}). First, it maturates and develops a phenotype $\phi$ that leads it to either survive or die; the probability to develop  $\phi$ given $\g$ is described by $D_\sn(\phi|\gamma)$ and the probability to survive given $\phi$ and the state $x_t$ of the environment by $S(\checkmark|\phi,x_t)$. Second, it reproduces into $\xi\geq 0$ offsprings, each with a genotype $\g'$ correlated to the genotype $\g$ of its parent; the probability to produce $\xi$ offsprings is described by $R_\sn(\xi)$ and the probability of $\g'$ given $\g$ by $H_\sn(\g'|\g)$. Individuals die after reproduction and the next generation consists of the newly born offsprings.

Assuming the population size to be large enough for stochastic effects (genetic drift) to be negligible, the number $N_{t}(\g)$ of individuals born at generation $t$ with genotypes $\g$ and the number $M_{t}(\g)$ of those reaching maturation satisfy the recursion
\bea
M_t(\g)&=&\int d\phi\ S(\checkmark|\phi)D_\sn(\phi|\g,x_t)N_{t}(\g)\label{eq:asex1}\\
N_{t+1}(\g')&=&k_\sn\int d\g\ H_\sn(\g'|\g) M_t(\g)
\eea
where $k_\sn=\int  d\xi R_\sn(\xi)\xi$ is the mean number of offspring per reproductive event.

\subsubsection{Monoecious sexual reproduction}

We first generalize to monoecious (hermaphroditic) sexual reproduction where each individual mate with a randomly chosen individual in the population to produce an average of $k_\sh$ offsprings:
\beq
M_{t}(\g)=\int d\phi\ S(\checkmark|\phi,x_t)D_{\sh}(\phi|\g)N_{t}(\g)\label{eq:mono1}
\eeq
\beq
N_{t+1}(\g')=k_\sh\int d\g_\sf d\g_\sm \ H_{\fm}(\g'|\g_\sf,\g_\sm)\frac{M_{t}(\g_\sm)}{M_{t}}M_{t}(\g_\sf)\label{eq:mono2}.
\eeq
Here $M_t=\int d\g M_t(\g)$ so that $M_t(\g_\sm)/M_t$ represents the probability for an individual with genotype $\g_\sf$ to mate with an individual with genotype $\g_\sm$. An offsprings inherits a genotype $\g'$ with probability $H_{\fm}(\g'|\g_\sf,\g_\sm)$, which depends a priori on the genotypes of the two parents, $\g_\sf$ and $\g_\sm$. Written in terms of the densities $n_t(\g)=N_{t}(\g)/\int d\g N_t(\g)$, this recursion is central to several previous models of population genetics~\cite{slatkin1970selection,roughgarden1972evolution,calvez2019asymptotic} and, more recently, to physical models of self-propelled particles with alignment interactions~\cite{degond2014local}.

When comparing to the asexual case, we will consider $k_\sh=k_\sn$, which assumes that each sexually reproducing individual produces as many offsprings as an asexually reproducing individual. A difference between $k_\sh$ and $k_\sn$ may, however, be justified by the presence of intrinsic costs to sexual reproduction, due for instance to the necessity to produce more gametes or to the difficulty of finding a mate. To illustrate how our conclusions generalize to $k_\sh<k_\sn$, we will also report results for $k_\sh=k_\sn/2$, which can be interpreted as a cost dominated by the number of gametes that each individual must produce, which is doubled in the sexual case. 

\subsubsection{Dioecious sexual reproduction}

To extend the formalism to dioecious sexual reproduction where two sexes are present, we assume that fecundity is limited by the number of females which choose their mate at random within the population of mature males. Assuming further that the sex of offsprings is chosen at random (sex-ratio $1/2$), this leads to the recursion
\beq
M_{\ss,t}(\g)=\frac{1}{2}\int d\phi\ S(\checkmark|\phi,x_t)D_{\ss}(\phi|\g)N_{t}(\g) \quad (\ss=\sf,\sm)\label{eq:dio1}
\eeq
\beq
N_{t+1}(\g')=k_\fm\int d\g_\sf d\g_\sm\ H_{\fm}(\g'|\g_\sf,\g_\sm)\frac{M_{\sm,t}(\g_\sm)}{M_{\sm,t}}M_{\sf,t}(\g_\sf)\label{eq:dio3}
\eeq
where $N_t(\g)$ is as before the number of newly born individuals with genotype $\g$ at generation $t$, and $M_{\sf,t}(\g)$ and $M_{\sm,t}(\g)$ are, respectively, the number of mature females and males with genotype $\g$, with $M_{\sm,t}=\int d\g_\sm M_{\sm,t}(\g_\sm)$ reporting the total number of maturate males. The ratio $M_{\sm,t}(\g_\sm)/M_{\sm,t}$ thus corresponds to the probability for a male with genotype $\g_\sm$ to be chosen by a female at generation $t$. We assume here that females and males are subject to the same selective pressure $S(\checkmark|\phi,x_t)$ but allow them to have different developmental modes $D_{\sf}(\phi|\g)$ and $D_{\sm}(\phi|\g)$. Finally, the probability $H_{\fm}(\g'|\g_\sf,\g_\sm)$ for an offspring to inherit a genotype $\g'$ depends a priori on the two genotypes of the parents, $\g_\sf$ and $\g_\sm$. We will take by default $k_\fm=k_\sh=k_\sn$. For monomorphic sexes ($D_\sf=D_\sm$), Eqs.~\eqref{eq:dio1}-\eqref{eq:dio3} are equivalent to Eqs.~\eqref{eq:mono1}-\eqref{eq:mono2} with $k_\sh=k_\sn/2$. For dimorphic sexes ($D_\sf\neq D_\sm$), on the other hand, the recursion defined by Eqs.~\eqref{eq:dio1}-\eqref{eq:dio3} has, to our knowledge, not been previously studied. {\col In the following, we analyze this recursion in the context of quantitative traits where the different kernels are Gaussian but note that the formalism is general and can also be applied to discrete traits and other kernels.}

\subsection{Questions}

The different modes of reproduction define different population dynamics: does it lead one mode of reproduction to be favored by natural selection? More specifically, as the answer may depend on the genetic and environmental constraints to which the populations are subject, we ask the following two questions:

Q1: Under what conditions is sexual reproduction advantageous over asexual reproduction?

Q2: Under what conditions is sexual dimorphism advantageous in dioecious populations?\\
The first question has been extensively studied, although no consensual solution has emerged~\cite{Otto:2002cn}. The second question, on the other hand, has to our knowledge not been examined mathematically. A particular challenge is known as the two-fold cost of males: in dioecious populations, males constitute half of the population but do not contribute directly to fecundity, in contrast to asexual or monoecious sexual populations where every individual can potentially contribute. This corresponds formally to the presence of a factor $1/2$ in Eq.~\eqref{eq:dio1} compared to Eq.~\eqref{eq:asex1} or Eq.~\eqref{eq:mono1}. This problem is usually presented in the context of Q1, when comparing dioecious sexual reproduction with asexual reproduction~\cite{maynard1971origin} (assuming no intrinsic cost of sex, i.e., $k_\fm=k_\sn$). It is, however, even more acute in the context of Q2, when comparing dioecious sexual reproduction with monoecious sexual reproduction, as only a factor $1/2$ differentiates the dynamics of monomorphic dioecious populations from that of monoecious populations (assuming here $k_\fm=k_\sh$). 

Addressing these questions requires defining the ``conditions'' and the nature of the possible ``advantages'' to which Q1 and Q2 refer. To this end, we adopt a simple parametrization of the different components of the model and present specific criteria for comparing populations differing by their reproductive or developmental modes.

\subsection{Basic model}\label{sec:basic}

{\col We consider the central model of quantitative genetics, the infinitesimal model~\cite{lynch1998genetics}, where $\g$ is a quantitative trait influenced by a large number of genes. Through the central limit theorem, this justifies to treat mutational and developmental noise as additive white Gaussian noise. The infinitesimal model played historically a major role in resolving the controversy between Mendelians and biometricians and continues today to be a corner-stone of evolutionary biology as well as a widely applied tool in plant and animal breeding~\cite{roff2007centennial}. As it is amenable to analytical calculations in the context of varying environments~\cite{Charlesworth:1993fg,Rivoire:2014kt,rivoire2016informations}, it provides particularly insightful results. We arrive at this model by making the following assumptions:}


$(i)$ we describe development from a genotype (breeding value) $\g\in\R$ to a phenotype $\phi\in\R$ by the addition of normally distributed noise, $\phi=\g+\nu$ with $\nu\sim\N(\s_{D}^2)$, where $\nu\sim\N(\s_{D}^2)$ indicates that $\nu$ is drawn from a normal distribution with zero mean and variance $\s_{D}^2$. The variance $\s_D^2$, which is sometimes referred to as the microenvironmental variance in quantitative genetics, is here called the developmental variance to distinguish it from the (macro)environmental variance $\s_E^2$ introduced below. When it depends on the mode of reproduction and on the sex, we also denote it by $\s_{D,\ss}^2$ with $\ss=\sn,\sh,\sf,\sm$ for, respectively, asexual, hermaphroditic, female and male individuals;

$(ii)$ we consider a stabilizing selection of the form $S(\checkmark|\phi,x_t)=e^{-(\phi-x_t)^2/(2\s_S^2)}$ where $x_t$ defines the optimal phenotype at generation $t$ and $\s_S^2$ the stringency of selection;

$(iii)$ for asexual reproduction, we assume that the genotype $\g'$ of an offspring is related to the genotype $\g$ of its parent by $\g'=\g+\nu$ with $\nu\sim\N(\s_M^2)$ where $\s_M^2$ is a mutational variance representing the effects of mutations; 

$(iv)$ for sexual reproduction, we assume that the genotype $\g'$ of an offspring is related to the genotypes $\g_\sf$ and $\g_\sm$ of its parents by $\g'=(\g_\sf+\g_\sm)/2+\nu$ with $\nu\sim\N(\s_M^2+\s_R^2)$ where $\s_R^2$ is a segregational variance that accounts for the variation introduced by recombination, in addition to the variation introduced by mutations.

Assuming a model for the process by which mutations and recombination operate on alleles, the genetic parameters $\s_M^2$ and $\s_R^2$ can be expressed in terms of more elementary parameters (number of loci, mutation rate, \dots)~\cite{burger2000mathematical}. Here, we do not make any assumption on the underlying mechanisms and treat $\s_M^2$ and $\s_R^2$ as fixed and independent parameters. As we will show that genetic variances are asymptotically constant, this is formally equivalent to assuming that populations have fixed genetic variances{\col, a key assumption of the infinitesimal model of quantitative genetics. This is clearly a strong assumption which requires particular conditions to be justified mathematically~\cite{barton2017infinitesimal} and whose applicability generally needs to be assessed through numerical simulations~\cite{Burger:1999wy,Waxman:1999vj}. We adopt it here but differ from the common practice in quantitative genetics by} parametrizing the model by $\s_M^2$ and $\s_R^2$ instead of the associated genetic variances $\s_\sn^2={\rm Var}(\g_\sn)$ and $\s_\sh^2={\rm Var}(\g_\sh)$. {\col As we show, this leads to} a more general and transparent interpretation of the results.

If $G_{\s^2}(x)=e^{-x^2/(2\s^2)}/\sqrt{2\pi\s^2}$ denotes a generic Gaussian function, we thus make the following assumptions:
\bea
(i)&D_\ss(\phi|\g)= G_{\s^2_{D,\ss}}(\phi-\g) \\
(ii)&S(\checkmark|\phi,x_t)= (2\pi\s_S^2)^{1/2}G_{\s^2_S}(\phi-x_t) \\
(iii)&H_\sn(\g'|\g)= G_{\s^2_M}(\g'-\g) \label{eq:Hsn}\\
(iv)&H_\fm(\g'|\g_\sf,\g_\sm)=G_{\s^2_M+\s_R^2}\left(\g'-\frac{1}{2}(\g_\sf+\g_\sm)\right) \label{eq:Hfm}
\eea
 where $\ss$ stands for either $\sn$, $\sh$, $\sf$ or $\sm$. Additionally, the initial distribution of genotypes $N_0(\g)$ is also assumed to be Gaussian, which is sufficient to ensure that it remains Gaussian at any subsequent time $t>0$.

$(v)$ Finally, we assume that the environment follows an autoregressive process
\beq
x_t=ax_{t-1}+b_t,\quad b_t\sim\N((1-a^2)\s_E^2),
\eeq
or, equivalently,
\beq
P(x_t|x_{t-1})=G_{(1-a^2)\s_E^2}(x_t-ax_{t-1}).
\eeq
As $\E[x_tx_{t+\tau}]=\s_E^2 a^\tau$, the parameter $a\in [0,1[$ encodes the time scale of the environmental fluctuations {\col ($\tau_E=-1/\ln a$)} which is to be compared with the generation time {\col $(\tau=1)$}, corresponding to $a=e^{-1}\simeq 0.36$. The parameter $\s_E^2\in [0,\infty[$, on the other hand, encodes the variance of the fluctuations, which is to be compared with the stringency of selection $\s_S^2$; an environmental change that kills 50\% of a previously perfectly adapted population thus corresponds to $\s^2_E/\s^2_S=2\ln 2\simeq 1.4$.  The environmental variance $\s_E^2$ corresponds to the scale of environmental fluctuations over multiple generations while $(1-a^2)\s_E^2$ is the scale of these fluctuations over a single generation.

We will examine several extensions of this basic model to account for the evolution of heredity and development, for the presence of spatial heterogeneities or of a mutational bias, and for the possibility of plasticity. We will also examine the case of directional selection where $x_t=ct+b_t$ with $c\neq 0$ and $b_t\sim\N(\s_E^2)$. Other environmental processes, such as random walks or periodic processes, can also be examined in a similar way~\cite{Charlesworth:1993fg}.

\subsection{Levels of analysis}

In the simplest instance of the previous model, the conditions to which the populations are subject comprise genetic constraints, parametrized by $\s_M^2,\s_R^2$, and environmental constraints, parametrized by $a,\s_E^2$ {\col and $\sigma_S^2$}. When now considering the evolutionary advantage that different modes of development or reproduction may confer, it is useful to distinguish two levels of analysis:

L1: What is optimal for populations in the long-term?

L2: What may effectively evolve under natural selection?\\
What is optimal at a population level may indeed not possibly or effectively result from natural selection.

A definition of optimality at the population-level (L1) is provided by the long-term growth rate, which is formally defined in the limit of an infinite number of generations by $\L=\lim_{t\to\infty}(1/t)\ln N_t$ where $N_t$ is the total population size at generation $t$~\cite{Rivoire:2011fy}. This growth rate corresponds to the geometric mean of the instantaneous fitness, $\L=\E[\ln N_t/N_{t-1}]$, which is well-known to be the relevant quantity for large populations in varying environments~\cite{lewontin1969population}. {\col Crucially, it is the quantity that quantifies the value of information in the context of evolutionary dynamics~\cite{Rivoire:2011fy}.} In the long term ($t\to\infty$), populations following our models will either become extinct or grow exponentially at a rate $\L_\ss$, with $\ss=\sn,\sh$ or $\fm$ depending on the mode of reproduction. The population with largest $\L_\ss$ is then considered optimal. We shall verify that this criterion describes the outcome of competitions between finite-size populations over a finite number of generations when they are subject to a common total carrying capacity. In such cases, the population that become fixed is most likely the population with largest growth rate $\L_\ss$. Formally, the two questions Q1 and Q2 at level L1 therefore amounts to comparing the growth rates of different populations subject to the same constraints but differing by their mode of reproduction (Q1) or their mode of development (Q2):

Q1/L1: What is the sign of $\L_\sh-\L_\sn$ as a function of $\s_M^2,\s_R^2$ and $a,\s_E^2{\col ,\s_S^2}$?

Q2/L1: What values of the developmental variances $\s^2_{D,\sf}$, $\s^2_{D,\sm}$ optimize $\L_\fm$ as a function of these same parameters?\\
In particular, the two-fold cost of males is overcome if $\L_\sh-\L_\sn>\ln 2$ in the context of Q1 (as $\L_\fm=\L_\sh-\ln 2$ if sexes are monomorphic) and if $\L_\fm-\L_\sh>0$ in the context of Q2 (which necessarily requires dimorphism). {\col More generally, $\L_\sh-\L_\sn$ quantifies the value of sexual reproduction over asexual reproduction and $\L_\fm-\L_\sh$ the value of dioecy over monoecy when comparing alternative ways to transmit information between generations.}

Addressing the two questions at level L2 requires us to augment the model with genotypic variables that are also subject to mutations but control either the mode of reproduction (Q1) or the mode of development (Q2), also known as modifier genes~\cite{karlin1974towards}. Formally,

Q1/L2: What is the long-term dynamics of a gene $\psi$ that controls the probability for an individual to reproduce sexually, as a function of $\s_M^2,\s_R^2$ and $a,\s_E^2{\col ,\s_S^2}$?

Q1/L2: What are the long-term dynamics of sex-specific genes $\d^\sf$ and $\d^\sm$ that control the developmental variances $\s_{D,\sf}^2$ and $\s_{D,\sm}^2$ as a function of these same parameters?

\section{Results}

We obtain analytical formulae for the long-term growth rates $\L_\sn,\L_\sh,\L_\fm$ as a function of the different parameters (Appendix~\ref{sec:formulae}). These formula recapitulate and extend previous results~\cite{Charlesworth:1993fg,Rivoire:2014kt}. {\col The formulae depend on the variances $\s_D^2,\s_E^2,\s_M^2,\s_R^2$ only via the ratios $\s_D^2/\s_S^2,\s_E^2/\s_S^2,\s_M^2/\s_S^2,\s_R^2/\s_S^2$, which implies that }we can set the stringency of selection to one ($\s_S^2=1$) without loss of generality. They also show that differences in growth rates are independent on the mean number of offsprings when considering $k_\sn=k_\sh=k_\fm$, and we take this number to be $k_\sn=k_\sh=k_\fm=2$ in the numerical simulations. 

\subsection{Genetic constraints and optimization}

In contrast to the environmental parameters, the genetic parameters $\s_M^2$, $\s_R^2$ are potentially subject to evolution. Optimizing the growth rates over these two parameters, we find that sexual reproduction neither provides an advantage nor a disadvantage over asexual reproduction (Appendix~\ref{sec:optima}). From this standpoint, sex can be adaptive only in the presence of genetic constraints. In the following, we therefore treat $\s_M^2$ and $\s_R^2$ as given genetic constraints, in addition to the environmental constraints. We will find, however, that the optimal value of the mutational variance for asexual populations, which we denote $\hat \s_M^2$, plays a particular role when comparing sexual and asexual reproduction.

\subsection{Sexual versus asexual reproduction: optimality}

\begin{figure}[t]
\begin{center}
\includegraphics[width=.92\linewidth]{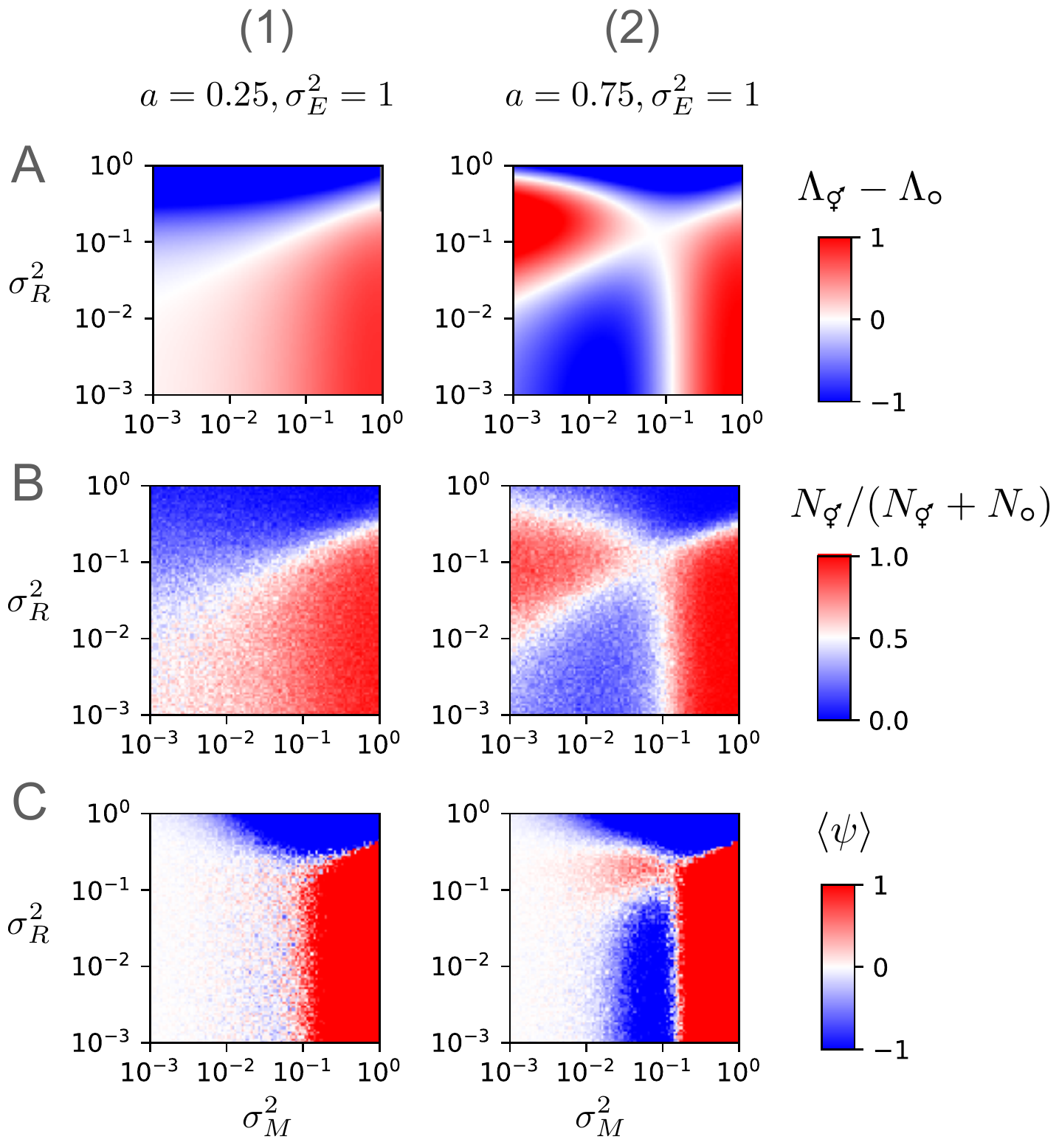}
\caption{Genetic conditions under which monoecious sexual reproduction is advantageous over asexual reproduction for {\col two environments differing by the time scale $\tau_E=-1/\ln a$ of their fluctuations}. {\bf A.} Difference $\L_\sh-\L_\sn$ between the growth rates of sexual and asexual populations. Values of $(\s_M^2,\s_R^2)$ in red represent conditions under which sex is advantageous {\col (see Appendix~\ref{app:mutload} for the contribution of the mutational load).} {\bf B.}~Fraction $N_\sh/(N_\sh+N_\sn)$ of sexually reproducing individuals after $T=250$ generations when starting from an initial population with an equal proportion of sexually and asexually reproducing individuals ($N_\sh=N_\sn$ at $t=0$). The simulations are performed under conditions where the total population size is maintained to $N=N_\sh+N_\sn=250$ and where each mature individual has $k=2$ offsprings, and the results correspond to averages over 100 independent simulations (Appendix~\ref{sec:simu}). {\bf C.} Mean value of the modifier gene $\psi$ that controls the probability to reproduce sexually in a model where this probability is subject to evolution, following Eq.~\eqref{eq:sexasexmod}. The results are averages over 100 replicate simulations starting from $N=250$ individuals with $\g=0$ and $\psi=0$, and ending after $t=250$ generations with a total population size maintained to $N=250$. For the smallest values of $\s_M^2$ and $\s_R^2$, no difference is observed (in white) and evolving towards one mode of reproduction or the other requires a larger number of generations (Appendix~\ref{app:evo_TM}).
Although A,B,C represent different quantities obtained under different assumptions, a remarkable overlap is apparent between the conditions under which sexual reproduction is optimal in the long-run (A), over-competes asexually reproducing populations after a finite number of generations (B) and evolves (C).
\label{fig:AH_opt}}
\end{center} 
\end{figure}

{\col As $\L_\sh-\L_\sn$ quantifies the value of sexual monoecious reproduction over asexual reproduction, the sign of $\L_\sh-\L_\sn$ indicates the environmental and mutational conditions under which sexual monoecious reproduction confers a long-term evolutionary advantage over asexual reproduction.} Here we assume $\s_D^2=0$ and discuss in Sec.~\ref{sec:space} how a finite developmental variance ($\s_{D,\sn}^2=\s_{D,\sh}^2>0$) changes quantitatively but not qualitatively the results. We are then left with 4 parameters, $\s_M^2$, $\s_R^2$, for the genetic constraints, and $a$, $\s_E^2$ for the environmental constraints. Displaying $\L_\sh-\L_\sn$ as a function of $\s_M^2,\s_R^2$ for two representative values of $a,\s_E^2$: (1)~$a=0.25$, $\s_E^2=1$ and (2)~$a=0.75$, $\s_E^2=1$  (Fig.~\ref{fig:AH_opt}A) reveals particular genetic and environmental conditions that must be satisfied for sexual reproduction to be advantageous over asexual reproduction ($\L_\sh>\L_\sn$ in red in Fig.~\ref{fig:AH_opt}A).

These conditions apply beyond the assumptions of infinite population size and infinite number of generations that underlie the calculations of the growth rates: they also decide the outcome of a competition between asexually and sexually reproducing populations after a finite number of generations when the total population size is subject to an upper bound. Starting from an equal mixture of asexual and sexual individuals, numerical simulations (Appendix~\ref{sec:simu}) indeed show that the mode of reproduction that becomes fixed is the one with largest growth rate (Fig.~\ref{fig:AH_opt}B).

\begin{figure}[t]
\begin{center}
\includegraphics[width=\linewidth]{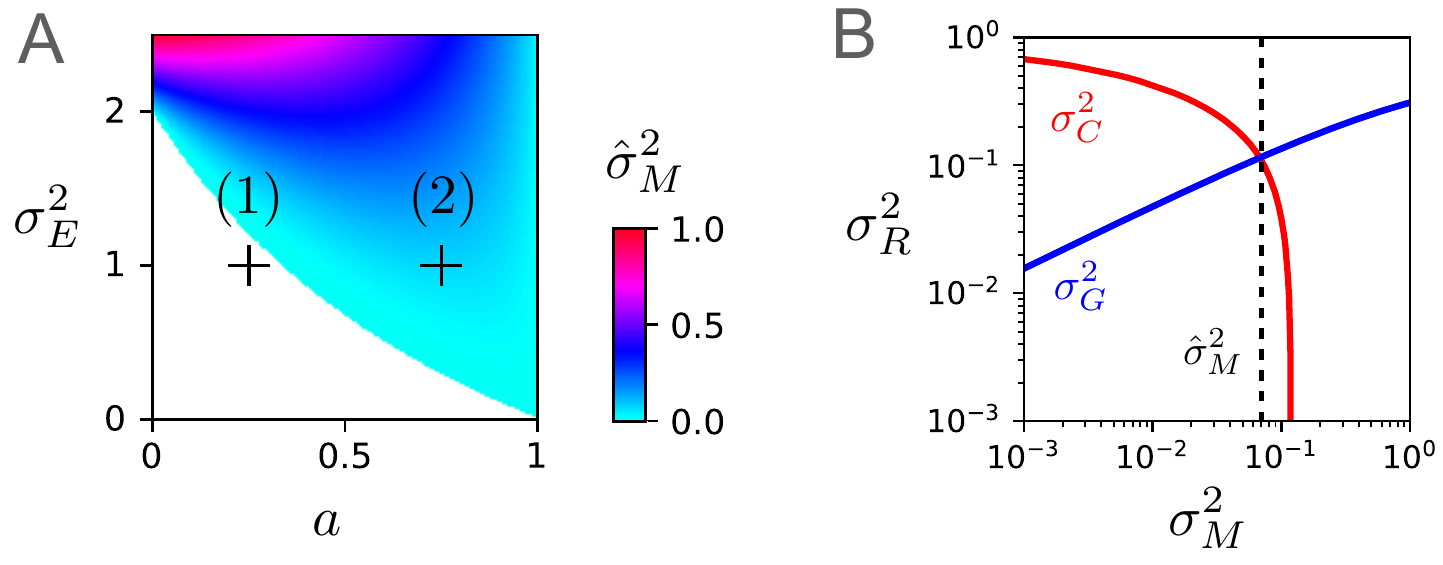}
\caption{{\bf A.} Optimal mutational variance $\hat\s_M^2$ for asexually reproducing populations as a function of $(a,\s_E^2)$. The conditions (1) and (2) of Fig.~\ref{fig:AH_opt} are indicated by crosses. {\col When $\s_E^2<2(1-a)/(1 + a)$, we have $\hat\s_M^2=0$ (white region). This condition corresponds to environmental fluctuations of sufficiently small amplitude $\s_E^2$ but also, less intuitively, of sufficiently small temporal correlations $\tau_E=-1/\ln a$. Informally, trying to keep up with a rapidly varying environment through random variations incurs a large mutational load that can make preferable the maintenance of a fixed trait $\g=0$ around which the optimal trait fluctuates.}
{\bf B.}~Thresholds $\s_G^2$ (in blue) and $\s_C^2$ (in red) separating the different regimes, here represented for the environmental conditions (2) of Fig.~\ref{fig:AH_opt}. The threshold $\s_G^2$ is, however, independent of the environmental conditions. The two curves meet at the value $\hat\s_M^2$ of $\s_M^2$ represented in A (black dotted line). When $\hat\s_M^2=0$, the threshold $\s_C^2$ in red is effectively absent, as illustrated by the environmental conditions (1) of Fig.~\ref{fig:AH_opt} {\col (see also Appendix~\ref{app:s2S} for the dependence on the stringency of selection $\sigma^2_S$).}
\label{fig:sM2}}
\end{center} 
\end{figure}

How to make sense of the phase diagrams of Fig.~\ref{fig:AH_opt}A? As a function of $\s_M^2,\s_R^2$, we have in the most general case 4 regimes, separated by two threshold functions $\s_R^2(\s_M^2)$ at which $\L_\sh=\L_\sn$ (Fig.~\ref{fig:sM2}B). Remarkably, one of these threshold functions, which we denote $\s_G^2$, is independent of the environmental conditions and given by (Appendix \ref{sec:sG2})
\beq\label{eq:s2G}
\s_G^2=\frac{\s_M^2}{4}\left[\sqrt{1+4\frac{\s_S^2+\s_D^2}{\s_M^2}}-1\right].
\eeq
This value has a simple interpretation: it is the value of the segregation variance $\s_R^2$ for which the variance of trait $\g$ in the population (the genetic variance) is the same for asexually and sexually reproducing populations (Appendix \ref{sec:sG2}). It also corresponds to the value of the segregation variance $\s_R^2$ under a Gaussian allelic approximation where $\g$ arises from a large number of alleles, $\g=\sum_{i=1}^L \g_i$, with each $\g_i$ assumed to be normally distributed in the population~\cite{burger2000mathematical} (Appendix \ref{sec:GAA}); in this model, the equivalence between genetic variances of asexually and sexually reproducing populations is in fact valid at any generation, beyond any steady-state assumption (Appendix \ref{sec:GAA}).

An equivalence of genetic variances is a sufficient but non necessary condition for the growth rates $\L_\sn$ and $\L_\sh$ to be the same. The second threshold function $\s_C^2$ corresponds to another solution, which exists only under some conditions, including condition (2) of Fig.~\ref{fig:AH_opt} but not condition (1). It crosses $\s_G^2$ at a particular value $\hat\s_M^2$ of $\s_M^2$ that has a simple interpretation: it is the value of $\s_M^2$ that optimizes the growth rate of asexual populations under the same environmental conditions. When $\s_R^2<\s_G^2$, $\s_C^2$ sharply decreases at a value of $\s_M^2$ only slightly above $\hat\s_M^2$. This may be interpreted as follows: sexual reproduction effectively decreases the mutational variance $\s_M^2$, which is beneficial when $\s_M^2>\hat\s_M^2$ but detrimental when $\s_M^2<\hat\s_M^2$. In the limit $\s_R^2\to 0$, this reduction of variance may be interpreted as a form of blending inheritance. The value of $\hat\s_M^2$ depends on environmental parameters (Fig.~\ref{fig:sM2}A). Remarkably, for sufficiently moderately varying environments, defined by $\s_E^2<2(1-a)/(1 + a)$, we have $\hat\s_M^2=0$ and sexual reproduction is therefore advantageous whenever $\s_R^2<\s_G^2$, irrespective of the value of $\s_M^2$. This is illustrated by condition (1) of Fig.~\ref{fig:AH_opt}, which is thus representative of a large class of environmental conditions.

{\col Our results are directly comparable to those of Charlesworth~\cite{Charlesworth:1993fg} who analyzed an equivalent model using a different parametrization and under the assumption that the genetic variance of sexual populations is larger, which corresponds to restricting to $\s_R^2>\s_G^2$.}

We assumed so far $k_\sh=k_\sn$. Taking instead $k_\sh=k_\sn/2$, which corresponds to an intrinsic two-fold cost of sex, the conditions for sexual reproduction to be advantageous over asexual reproduction are more stringent. We find that sex can be favored only in one of the four regimes described defined in Fig.~\ref{fig:sM2}B: when $\s_G^2<\s_R^2<\s_C^2$, provided $\s_E^2$ is sufficiently large (Appendix~\ref{app:2fold}). As populations are prone to extinction in largely varying environments, an additional condition is that each female produces in average a sufficiently large number $k_\sh$ of offsprings (Appendix~\ref{app:2fmore}).

\subsection{Sexual versus asexual reproduction: evolution}

Optimality of a trait at the population level is generally not sufficient to ensure that it may effectively evolve. To study this question, we generalize our basic model to include a genetic factor $\psi$ that controls how individuals reproduce. Specifically, we assume that an individual with genotype $\psi$ reproduces sexually with a probability $P(\psi)$ by mating with a randomly chosen individual, and asexually otherwise. We further assume the modifier gene $\psi$ to be transmitted through the females and subject to the same mutational variance $\s_M^2$ as $\g$. This corresponds to leaving Eq.~\eqref{eq:mono1} unchanged but replacing Eq.~\eqref{eq:mono2} by
\begin{widetext}
\beq\label{eq:sexasexmod}
N_{t+1}(\g',\psi')=k\int d\g_\sf d\psi_\sf \ \Bigg[
P(\psi_\sf)\int d\g_\sm   d\psi_\sm H_{\fm}(\g'|\g_\sf,\g_\sm)\frac{M_{t}(\g_\sm,\psi_\sm)}{M_{t}}+(1-P(\psi_\sf))H_{\sn}(\g'|\g_\sf)\Bigg]H_{\sn}(\psi'|\psi_\sf)M_{t}(\g_\sf,\psi_\sf).
\eeq
\end{widetext}
We take $P(\psi)=1/(1+e^{-\psi})$ so as to map $\psi\in\R$ into $P(\psi)\in [0,1]$ through a simple monotonic function that permits the evolution towards $P(\psi)\simeq 0$ or $P(\psi)\simeq 1$ when $\psi$ takes large absolute values. The results indicate that sexual reproduction typically evolves whenever sexual reproduction is advantageous (Fig.~\ref{fig:AH_opt}C). The conclusions derived from a comparison between long-term growth rates can therefore be obtained as the result of an evolutionary process.

\subsection{Dioecy and sexual dimorphism: optimality}

\begin{figure}[b]
\begin{center}
\includegraphics[width=\linewidth]{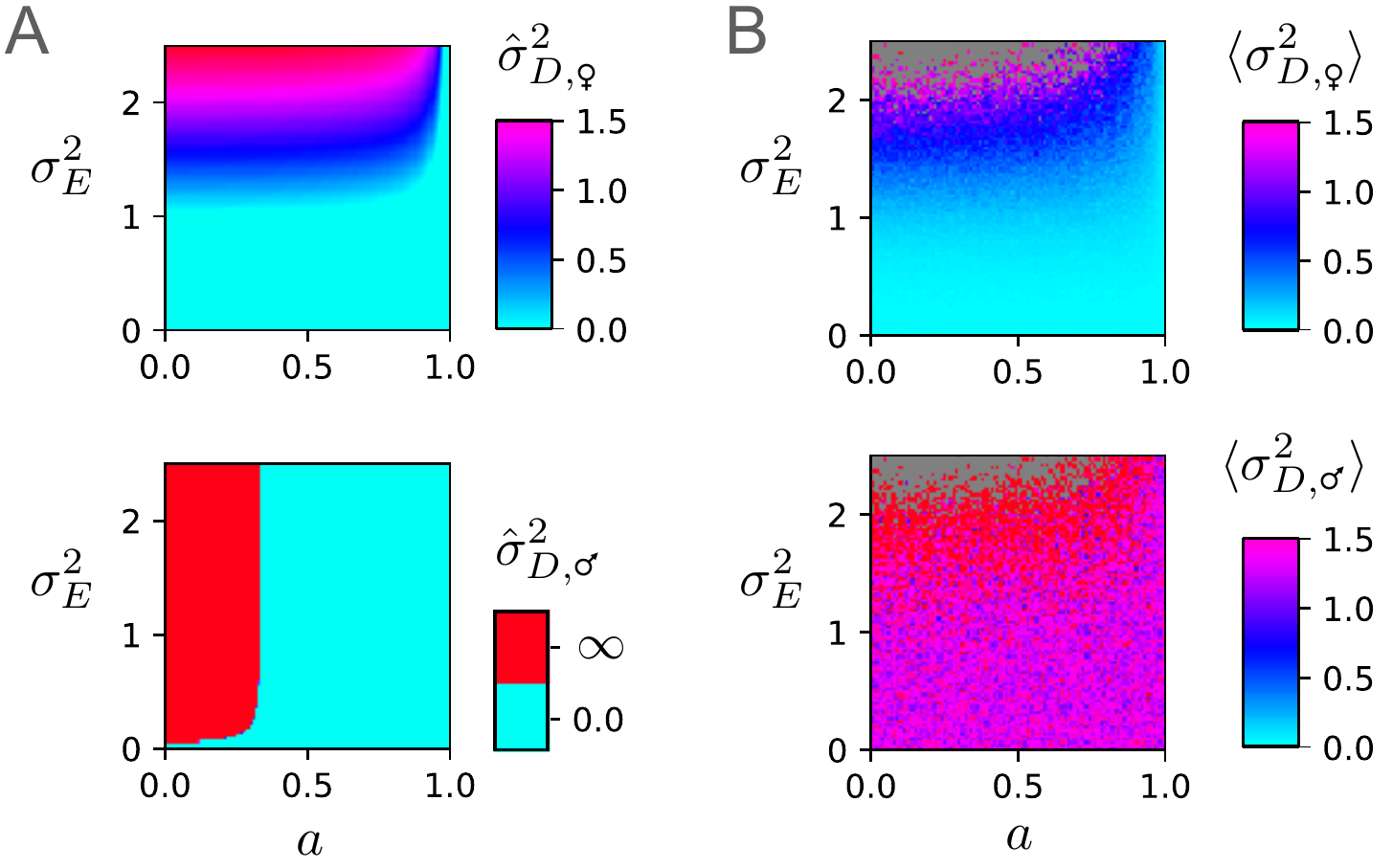}
\caption{{\bf A.} Optimal female and male developmental variances as a function of environmental conditions $(a,\s_E^2)$ for dioecious populations with $\s_M^2+\s_R^2=10^{-2}$ . In the limit $\s_M^2+\s_R^2\to 0$, $\hat\s_{D,\sf}^2$ is non-zero when $\s_E^2>\s_S^2$ (with here $\s_S^2=1$) and $\hat\s_{D,\sm}^2$ when $a<1/3$, in which case it is in fact infinite. {\bf B.} Evolutionary results of numerical simulations over $T=250$ generations with populations of size $N=250$ where the developmental variances are subject to evolution, following Eq.~\eqref{eq:evosex}. The gray points for large values of $\s_E^2$ correspond to cases where more than 10\% of the 100 populations that were independently simulated became extinct. The results for $\langle\s_{D,\sm}^2\rangle$ reflect here the initial conditions, which correspond to $\s_{D,\ss}^2=1$ (Appendix~\ref{app:evomore}). They are also contingent to the mode of transmission of the modifier genes (Appendix~\ref{app:dimore}).\label{fig:dimo}}
\end{center} 
\end{figure}

Dioecy opens the possibility of an asymmetry between the sexes. In our model, the two sexes share a common distribution of genotypes $\g$ but may display different phenotypes as a result of different developmental variances $\s_{D,\sf}^2$ and $\s_{D,\sm}^2$. To investigate whether a phenotypic dimorphism may confer an evolutionary advantage, we optimize the long-term growth rate $\L_{\fm}$ over the two sex-specific developmental variances $\s_{D,\sf}^2$ and $\s_{D,\sm}^2$ in a context where all other parameters are fixed. We find that environmental conditions indeed exists under which a dimorphism is advantageous (Fig.~\ref{fig:dimo}A). These conditions take a particularly simple form in the limit of small genetic variations, $\s_M^2+\s_R^2\ll \s_S^2$, where a non-zero female developmental variance $\hat\s_{D,\sf}^2=\s_E^2- \s_S^2$ is advantageous when the variance of the environmental fluctuations is large ($\s_E^2> \s_S^2$) while a non-zero male developmental variance $\s_{D,\sm}^2=\infty$ is advantageous when the timescale of the environmental fluctuations is short ($a< 1/3$). 

The selective pressure on the two developmental variances is, however, on different scales as $\L_\fm(\s_{D,\sf}^2,\s_{D,\sm}^2)\simeq \L_0(\s_{D,\sf}^2)+\L_1(\s_{D,\sf}^2,\s_{D,\sm}^2)(\s_M^2+\s_R^2)/\s_S^2$ when $\s_M^2+\s_R^2\ll\s_S^2$ (Appendix~\ref{sec:sh0}). Consequently, the selective pressure on male developmental variances is much weaker than the selective pressure on female developmental variances. Besides, an infinite developmental variance ($\hat\s_{D,\sm}^2=\infty$) is conceivable only in populations of infinite size. With finite populations, an upper bound on $\s_{D,\sm}^2$ arises from the need to maintain a sufficient number $N_{\sm,\rm min}$ of surviving males at each generation, which takes the form $\s_{D,\sm}^2\ll N_{\sm,\rm min}^2$ when $\s_M^2+\s_R^2\ll\s_S^2$ (Appendix \ref{sec:finitepop}). {\col All together, an asymmetry between the two sexes is not only present in the optimal values of their developmental variances but also in the strength of the selective pressure to which these developmental variances are subject.}

Because the contribution of $\s_{D,\sm}^2$ {\col to the growth rate $\L_\fm(\s_{D,\sf}^2,\s_{D,\sm}^2)$} is of order $\s_M^2+\s_R^2$, the growth rate of dioecious populations with optimal developmental variances $\hat\s_{D,\sf}^2$ and $\hat\s_{D,\sm}^2$, does not exceed significantly the growth rate of monoecious populations with optimal developmental variance $\hat\s_{D,\sh}^2$ (Appendix~\ref{app:r}). {\col For the basic model introduced in Sec~\ref{sec:basic}, we therefore reach the conclusion that sexual dimorphism is not sufficient to overcome the two-fold cost of males that dioecious populations incur compared to monoecious populations. As we show below, this two-fold cost can be overcome when the model includes a mutational bias or directional selection (Sec.~\ref{sec:bias}).}

Finally, we may question the assumption that the segregation variance takes a fixed value $\s_R^2$ independent of the developmental variances. Under the Gaussian allelic model, for instance, $\s_R^2=\s_G^2$, and Eq.~\eqref{eq:s2G} can be generalized to show that $\s_G^2$ depends on $\s_{D,\sf}^2$ and $\s_{D,\sm}^2$ (Appendix~\ref{sec:GAA}). Repeating the analysis under this assumption leads, however, to similar results, indicating that the adaptive advantage of sexual dimorphism is robust to the exact form that the segregation variance $\s_R^2$ takes (Appendix~\ref{app:r}).

\subsection{Dioecy and sexual dimorphism: evolution}

To analyze whether sexual dimorphism may evolve despite the reserves that we made, we augment the model to include two modifier genes $\d^\sf$ and $\d^\sm$ that control developmental variances of each sex specifically. This corresponds to recursions of the form
\begin{widetext}
\bea\label{eq:evosex}
M_{\ss,t}(\g,\d^\sf,\d^\sm)&=&\frac{1}{2}\int d\phi S(\checkmark|\phi,x_t)D(\phi|\g,\d^\ss)N_{t}(\g,\d^\sf,\d^\sm)\qquad (\ss=\sf,\sm)\\
N_{t+1}(\g',{\d'}^\sf,{\d'}^\sm)&=&k \int \prod_{\ss=\sf,\sm} d\g_\ss d\d^\sf_\ss d\d^\sm_\ss  \ H({\d'}^\ss|\d_\sf^\ss,\d_\sm^\ss) H_{\fm}(\g'|\g_\sf,\g_\sm)\frac{M_{\sm,t}(\g_\sm,\d^\sf_\sm,\d^\sm_\sm)}{M_{\sm,t}}M_{\sf,t}(\g_\sf,\d^\sf_\sf,\d^\sm_\sf)\nonumber
\eea
\end{widetext}
where $D(\phi|\g,\d^\ss)=G_{e^{\d^\ss}}(\phi-\g)$, i.e., $\s_{D,\sf}^2=e^{\d^\sf}$ and $\s_{D,\sm}^2=e^{\d^\sm}$, a choice made to map $\d^\ss\in\R$ into $\s^2_{D,\ss}\in\R^+$ through a simple monotonic function, with the possibility to easily obtain $\s_{D,\ss}^2\simeq 0$ when $\d^\ss$ takes negative values.

Assuming that the modifiers are transmitted through the females, as we did in Eq.~\eqref{eq:sexasexmod}, corresponds to $H({\d'}^\ss|\d_\sf^\ss,\d_\sm^\ss)=H_\sn({\d'}^\ss|\d_\sf^\ss)$, in which case we indeed observe the evolution of sexual dimorphism (Fig.~\ref{fig:dimo}B). While the developmental variances of females reach values conform to the optima derived from the optimization of $\L_\fm$, this is not the case for the developmental variances of males, which are strongly dependent on initial conditions, consistent with a very weak selective pressure (Appendix~\ref{app:evomore}). Besides, the results depend on the mode by which the modifier genes are transmitted. Assuming that they are subject to recombination, $H({\d'}^\ss|\d_\sf^\ss,\d_\sm^\ss)=H_\fm({\d'}^\ss|\d_\sf^\ss,\d_\sm^\ss)$, or that they are separately inherited for each sex, $H({\d'}^\ss|\d_\sf^\ss,\d_\sm^\ss)=H_\sn({\d'}^\ss|\d_\ss^\ss)$, leads to monomorphic populations (Appendix~\ref{app:dimore}). We may interpret these results as a consequence of sexual selection counteracting selection at the population level. Transmitting modifiers through the females is indeed special in this respect, as males at one generation do not inherit any direct information from males of the previous generation.

\subsection{Developmental noise and spatial heterogeneities}\label{sec:space}

\begin{figure}[t]
\begin{center}
\includegraphics[width=\linewidth]{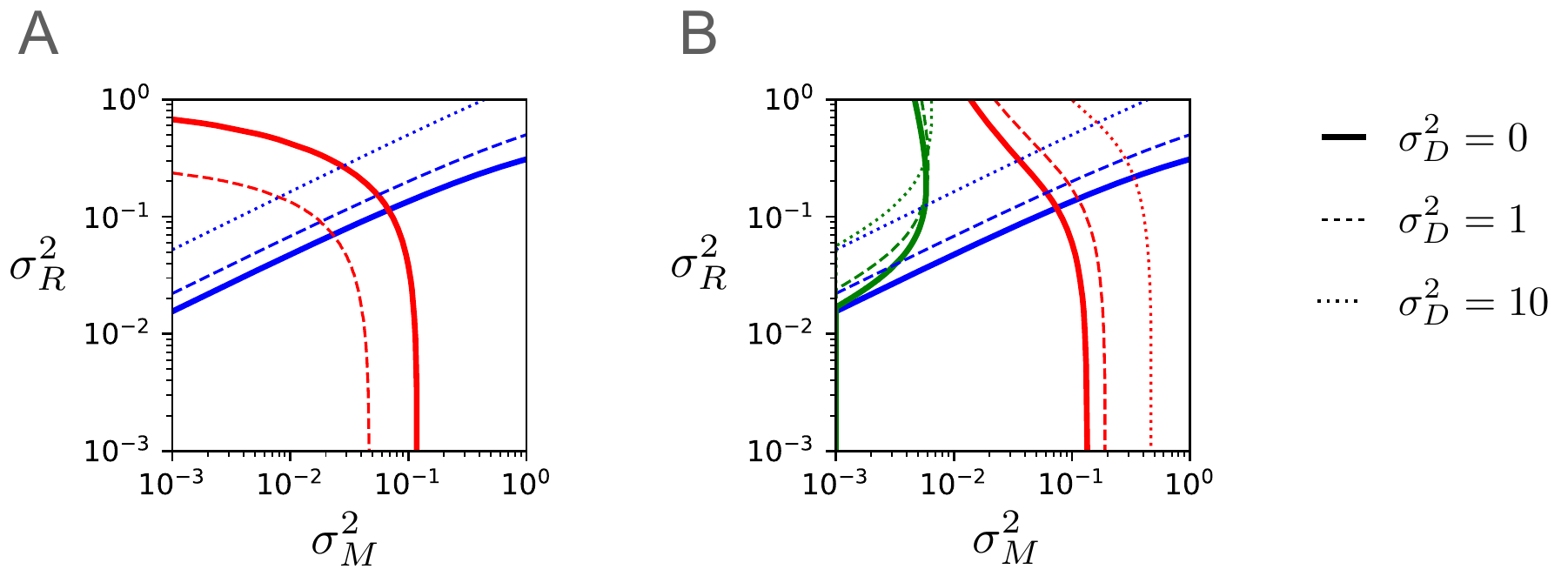}
\caption{{\bf A.} Extension of Fig.~\ref{fig:sM2}B to different values of developmental variances $\s_D^2$. The blue curve corresponds to $\s_G^2$, the value of the segregation variance $\s_R^2$ above which sexual populations have larger genetic variance than asexual populations. The red curve corresponds to $\s_C^2$, such that sex is advantageous when $\s_G^2<\s_R^2<\s_C^2$ or $\s_C^2<\s_R^2<\s_G^2$. Full lines are for $\s_D^2/\s_S^2=0$ as in Fig.~\ref{fig:sM2}B, dashed lines for $\s_D^2/\s_S^2=1$ and dotted lines for $\s_D^2/\s_S^2=10$. When $\s_D^2/\s_S^2=10$, we have $\s_C^2\to-\infty$ and the corresponding line is therefore absent, as in case (1) of Fig.~\ref{fig:AH_opt}. {\bf B.} Similar to A but for a directionally varying environment with $c=0.1$, $a=0$, $\s_E^2=0$. The curves for $\s_G^2$, which do not depend on the environment, are identical to A. The curves for $\s_C^2$, on the other hand, are shifted to larger values with increasing developmental variances $\s_D^2$. Additionally, there are now conditions for which a two-fold cost for sex is overcome ($\L_\sh>\L_\sn+\ln 2$), corresponding to values of $\s_M^2$ on the left side of the green curves.\label{fig:sigmaD}}
\end{center} 
\end{figure}

For simplicity, we compared so far sexual and asexual reproduction in the absence of developmental noise. Assuming instead a finite and common developmental variance $\s_{D,\sn}^2=\s_{D,\sh}^2$ shifts the boundaries between the different regimes (Fig.~\ref{fig:sigmaD}A): the value of $\s_G^2$ increases, as indicated by Eq.~\eqref{eq:s2G}, while the value of $\s_C^2$ decreases. 

A larger developmental variance may result from spatial environmental heterogeneities. For instance, we may consider that differences in local environments contribute to the developmental variance, $\s^2_{D,\rm tot}=\s^2_D+ \s^2_{D, \rm loc}$, or that different local environments are associated with different optimal phenotypes $y_t$ defining the selective pressure $S(\checkmark|\phi,y_t)$. If these locally optimal phenotypes $y_t$ are distributed normally around a mean optimal value $x_t$ with variance $\s^2_{E,\rm loc}$, the two points of view are equivalent and formally amount to redefining the developmental variance $\s_D^2$ by $\s_D^2+ \s^2_{D, \rm loc}+\s^2_{E,\rm loc}$ (Appendix~\ref{sec:spatial}). Whether spatial heterogeneities favor sex has thus no simple general answer but depends on the values of the genetic parameters $\s_M^2,\s_R^2$.

\subsection{Mutational biases and directional selection}\label{sec:bias}

We assumed mutations to be on average neutral but the model can be extended to analyze cases where their average effect is $c_M\neq 0$. This corresponds to generalizing $(iii)$ to $\g'=\g+\nu+c_M$ and $(iv)$ to $\g'=(\g_\sf+\g_\sm)/2+\nu+c_M$. Formally, this is equivalent to introducing a systematic drift $c_E$ in the environment, i.e., to generalize $(v)$ to $x_{t+1}=ax_t+b_t+c_Et$. The growth rates of the models that include $c_M$ and $c_E$ depend indeed on these parameters only via the combination $c=c_E-c_M$. The genetic or environmental origin of this particular constraint is therefore irrelevant.

 Models with $c\neq 0$ are in many respects similar to models with $c=0$ but a large value of $a$ (Fig.~\ref{fig:sigmaD} and Appendices~\ref{app:2fold}, \ref{app:2fmore}, \ref{app:r}, \ref{app:2bis}). This is not surprising, as the parameters $a>1/3$ and $c>0$ similarly induce cross-generation environmental correlations. Two differences are nevertheless worth mentioning: the two-fold cost of sex is overcome in a larger range of conditions (Appendix~\ref{app:2fold}) and larger developmental variances increase $\s_C^2$ (Fig.~\ref{fig:sigmaD}B). These results recapitulate the conclusions {\col of Charlesworth~\cite{Charlesworth:1993fg} who compared autocorrelated ($a>0, c=0$) and directed ($a=0,c>0$) environments in the regime $\s_R^2>\s_G^2$ and found that the two-fold cost of sex can be overcome only with directed environments.} Additionally, we find here that the two-fold cost of dioecy relative to monoecy can in principle be overcome under a sufficient mutational bias or/and directed selection (Appendices~\ref{app:2fold2} and \ref{app:S9}). {\col This corresponds, however, to situations where the mean number $k$ of offsprings per reproductive event must be sufficiently large for the population to escape extinction ($\Lambda>0$, Appendices~\ref{app:2fmore} and \ref{app:S9}). This motivates an extension of the model to include phenotypic plasticity, which defines a generic mechanism by which the probability of extinction can be reduced.} 

\subsection{Phenotypic plasticity}

One impediment to the evolution of dioecy through sexual dimorphism is the cost incurred by males, whose surviving fraction may be very small (Appendix~\ref{app:S9}). Phenotypic plasticity can alleviate this effect without comprising the benefit of sexual dimorphism at the population level. 

We assumed indeed that development from $\g$ to $\phi$ was independent of the environment but the model can be extended to include forms of phenotypic plasticity where $\phi$ also depends on $x_t$. For instance, we may consider that $\phi=(1-\k)\g+\k x_t+\nu$ with a reaction norm $\k\in [0,1]$ and, as before, a developmental noise $\nu\sim\N(\s_{D}^2)$. The absence of plasticity that we assumed so far corresponds to the particular case $\k=0$ (see Appendix~\ref{sec:plasticity} for a generalization to non-convex combinations of $\g$ and $x_t$). 

Growth rates are trivially optimized by taking $\k=1$ and $\s_{D}^2=0$, which effectively eliminates any effect of natural selection since $\phi=x_t$. Constraints are expected to prevent this optimum to be reached. One may for instance assume that $\phi=(1-\k)\g+\k y_t+\nu$ where the local environment $y_t$ experienced by an individual during development is only partially correlated to the selective pressure, e.g. $y_t=x_t+\nu$ with $\nu\sim\N(\s_L^2)$, or subject to a delay, e.g., $y_t=x_{t-\tau}$ with $\tau>0$. Growth rates can be obtained analytically in these cases but a simpler constraint is to assume that non-zero reaction norms $\k>0$ have a direct selective cost, which corresponds to multiplying $S(\checkmark|\phi,x_t)$ by a factor $C(\k)<1$ that is a decreasing function of $\k$~\cite{Chevin:2010cw}. This case can formally be mapped onto the basic model with an effective developmental variance that depends on $\k$ (Appendix~\ref{sec:plast_mapping}). Optimizing and evolving plasticity is therefore similar to optimizing and evolving developmental variances. For instance, in the dioecious case where the sexes may have different reaction norms $\k_\sf$ and $\k_\sm$, we find that in autoregressive environments $\k_\sm$ has no incidence on the growth rate while $\k_\sf$ effectively reduces $\s_E^2$ by a factor $(1-\k_\sf)^2$ (Appendix~\ref{sec:plasticity}). The optimal plasticity is then dimorphic, with $\hat\k_\sm$ taking arbitrary values and $\hat\k_\sf$ setting a balance between large values of $\k$ that minimize the effect of natural selection and small values of $\k$ that maximize $C(\k_\sf)$. Finally, we note that while plastic and non-plastic populations can be equivalent at the genotypic level, they are generally very different at the phenotypic level, and display in particular different phenotypic variances (Appendix~\ref{sec:phenononeq}). {\col In particular, plasticity} allows for higher survival during maturation, with no loss at all in the extreme limit of perfect plasticity.

\section{Discussion}

We studied a model of sexually reproducing population that generalizes previous models of information processing in asexual populations subject to varying environments~\cite{lachmann1996inheritance,bergstrom2004shannon,kussell2005phenotypic,Rivoire:2011fy,Rivoire:2014kt}. {\col The long-term growth rates that we calculate quantify the value of sexual reproduction and sexual dimorphism as schemes for transmitting information between generations. In particular, they} identify genetic and environmental conditions under which sexual reproduction and sexual dimorphism are optimal and may evolve. {\col The Gaussian model that we solve analytically corresponds to the infinitesimal model of quantitative genetics, which applies to complex traits under the influence of many genes. This model plays a fundamental role in population genetics~\cite{lynch1998genetics}, similarly to the Gaussian channel in information theory~\cite{cover1999elements}. Our general framework, however, is not restricted to this model and can also be applied to models with discrete traits.}

{\col In the Gaussian context, our comparison of sexual and asexual reproductions mirrors an analysis by Charlesworth~\cite{Charlesworth:1993fg}, who similarly studied} the environmental conditions under which sexual reproduction can be adaptive. Our results are consistent, showing that a steadily changing environment is most favorable. We differ, however, in our parametrization of the model and in our interpretation of some of the results. Charlesworth compared populations with given variance of the trait in the population (given genetic variances) assuming that sexual reproduction leads to higher genetic variance. We parametrize the mechanisms generating mutations and recombination by two more elementary parameters, a mutational variance $\s_M^2$ and a segregation variance $\s_R^2$, from which we derive both the genetic variance and the long-term growth rate (fitness) of the population. We find that sexual reproduction leads to higher genetic variance only for sufficiently large $\s_R^2$, namely $\s_R^2>\s_G^2$ where $\s_G^2$ is given by Eq.~\eqref{eq:s2G}, independently of environmental variations. As a function of the two genetic parameters $\s_M^2,\s_R^2$, we therefore obtain in the most general case four phases (Fig.~\ref{fig:AH_opt}), defined on one hand by whether sexual reproduction increases the genetic variance, which is independent of environmental conditions, and, on the other hand, by whether this increase is beneficial, which depends on the environmental conditions.

The advantage of sexual reproduction may thus be attributed to its ability to purge deleterious mutations either because it reduces variations, which can be beneficial when the mutational variance is too high, or because it increases them, which can be beneficial in presence of a mutational bias. The difference is significant: the first case is akin to the reduction of diversity attributed to blending inheritance while the second corresponds to recombination breaking down negative linkage disequilbrium~\cite{kondrashov1988deleterious,Otto:1998uq}. Similarly, varying environments may favor sexual reproduction either for providing more beneficial variations or for reducing detrimental variations. This second case is, maybe counter-intuitively, relevant when environments vary rapidly, as for instance in presence of co-evolving pathogens~\footnote{With auto-regressive environments the requirement is $\s_E^2<2(1-a)/(1+a)$, which includes drastic fluctuations that wipe out 90\% of a previously perfectly adapted population when $a=0$ and $\s_E^2=2$}.

An additional value of using the mutational variance $\s_M^2$ and the segregation variance $\s_R^2$ as parameters is the connection that it allows with the problem of optimal mutation rate in asexual populations, where the key parameter is $\s_M^2$. The optimal mutational variance $\hat\s_M^2$ in this problem, which depends on the fluctuations of the environment, defines indeed the point at which the four phases meet (Fig.~\ref{fig:sM2}). Defining parameters for the genetic mechanisms also leads us to notice that a mutational bias has formally the same implications as a directional bias. This is important as the presence of a systematic bias towards deleterious mutation may represent a more generic constraint than a steadily moving environment. The relevance of this constraint for the evolution of sex has, however, been only recently stressed~\cite{Vanhoenacker:2018cj}. Our approach also suggests that the opposition that is often made between constraints of environmental or genetical origin may be misguided, as constraints of same origin may be of very different nature (e.g., mutational variance versus mutational bias) while constraints of different origin may be of same nature (e.g., directional selection and mutational bias). A genetic constraint at the level of the mutational variance appears, however, essential for sexual reproduction to possibly confer any adaptive advantage.

Our model also formalizes and rigorously examines the possibility that sexual dimorphism may be adaptive in changing environments. This possibility was proposed by Geodakyan~\cite{Geodakyan:1965,Geodakyan:2015} but, to our knowledge, not previously examined mathematically. Under this hypothesis, females are more plastic or more subject to developmental noise than males, which permits an efficient integration of environmental information while preserving fecundity: the integration of information is performed by the males, whose phenotypes faithfully represent their genotypes while females are protected from direct elimination through selection by expressing phenotypes more loosely related to their genotypes. The environmental information obtained by males then flows to females in the next generation. This separation of roles in information processing has been asserted to be enough to overcome the two-fold cost of dioecy with respect to monoecy, thus providing an adaptive explanation for the ubiquitous presence of two sexes. {\col By quantifying the value of this information scheme,} our model shows that sexual dimorphism can indeed confer an adaptive advantage but that its evolution is subject to several limitations. Within our model, Geodakyan's scenario is therefore theoretically possible but only under specific conditions that make it unlikely to provide a generic explanation for the evolution of sexual dimorphism. Whether our conclusions hold in more realistic generalization of our model remains, however, to be examined.

While integrating different environmental and genetic constraints and accounting for some forms of spatial heterogeneities and phenotypic plasticity, our model indeed rests on strongly simplifying assumptions and cannot pretend to summarize the full range of factors that have been considered to play a role in the evolution of sex~\cite{Otto:1998uq,Otto:2002cn}. A strong assumption is that recombination can be described by a Gaussian model with fixed segregation variance $\s_R^2$. This assumption, which is known as the infinitesimal model, can be justified when the traits arise from the additive contribution of a large number of alleles, each contributing by an infinitesimal effect~\cite{barton2017infinitesimal}. Numerical simulations of models with a finite number of alleles show that 
sexual reproduction can lead to an increasing genetic variance~\cite{kondrashov1996high,Burger:1999wy,Waxman:1999vj}. While in contradiction with the infinitesimal model, these studies nevertheless concur in finding that directional selection, and therefore mutational biases, are favorable to sexual reproduction, as the underlying mechanism precisely rests on a larger genetic variance.

{\col By extending to sexual reproduction the quantification of biological information, our work invites an extension of the explanations of biological diversity~\cite{lachmann1996inheritance,Rivoire:2014kt,uller2015incomplete,mcnamara2016detection,mayer2016diversity} and the formal analogies~\cite{bergstrom2004shannon,kussell2005phenotypic,Rivoire:2011fy,Rivoire:2014kt,kussell2006polymer,skanata2016evolutionary,hirono2015jarzynski,kobayashi2015fluctuation,vinkler2016analogy,genthon2020fluctuation} previously developed for asexual populations. For instance, it would be interesting to generalize the formulation and interpretation of the models in terms of lineages to sexual populations whose genealogies are not tree-like~\cite{georgii2003supercritical,leibler2010individual,garcia2019linking}. Our work also motivates generalizations to account for other forms of horizontal transmission of information between individuals.}

\vspace{2cm}

\acknowledgements

We are grateful to Denis Roze for comments and suggestions of references. Funding for this work was partly provided by the Alexander von Humboldt Foundation in the framework of the Sofja Kovalevskaja Award endowed by the German Federal Ministry of Education and Research.

\appendix

\setcounter{section}{0} 

\section{Growth rates $\L_\sn,\L_\sh,\L_\fm$}\label{sec:formulae}

We consider here the basic model with an environment following $x_{t+1}=ax_t+b_t+ct$ with $b_t\sim\N((1-a^2)\s_E^2)$. 

\subsection{Analytical formulae}

The growth rates $\L_\ss$ for asexual ($\ss=\sn$), monoecious ($\ss=\sh$) and dioecious ($\ss=\fm$) reproduction involve the same function $L$ defined by
\bea
&&L(\alpha_\ss,\eta_\ss,a,\s_E^2/\s_S^2,c)=\frac{1}{2}\ln(\eta_\ss\a_\ss)\\
&&-\eta_\ss\a_\ss\left[\frac{(1-a)}{(1-a\a_\ss)(1+\a_\ss)}\frac{\s_E^2}{\s_S^2}+\frac{c^2}{2(1-a)^2(1-\a_\ss)^2}\right]\nonumber
\eea

The parameters $\a_\ss$ and $\eta_\ss$ are expressed in terms of the following variables, defined for $\ss=\sn,\sh,\sf,\sm$:
\beq
\eta_\ss=\frac{\s_S^2}{\s_S^2+\s_{D,\ss}^2},\qquad \beta_\ss=\frac{\s_{H,\ss}^2}{\s_S^2+\s_{D,\ss}^2}.
\eeq
where $\s_{H,\sn}^2=\s_M^2$ while $\s_{H,\ss}^2=\s_M^2+\s_R^2$ for $\ss=\sh,\sf,\sm$.

For asexual reproduction, $\L_\sn=\ln k+L(\alpha_\sn,\eta_\sn,a,\s_E^2/\s_S^2,c)$ with 
\beq\label{eq:asn}
\alpha_\sn=\frac{2+\beta_\sn-\sqrt{\beta_\sn(4+\beta_\sn)}}{2},\qquad 
\eeq

For monoecious reproduction, $\L_\sh=\ln k+L(\alpha_\sh,\eta_\sh,a,\s_E^2/\s_S^2,c)$ with 
\beq\label{eq:ash}
\alpha_\sh=\frac{3+2\beta_\sh-\sqrt{1+12\beta_\sh+4\beta_\sh^2}}{2},
\eeq

For dioecious reproduction, $\L_\fm=\ln (k/2)+L(\alpha_\fm,\eta_\fm,a,\s_E^2/\s_S^2,c)$ with 
\bea\label{eq:afm}
\a_\fm=\frac{1}{2}(\a_\sf+\a_\sm),\quad\a_\sf=\frac{1}{1+\b_\sf/(1-\a_\fm/2)},\nonumber\\
\quad\a_\sm=\frac{1}{1+\b_\sm/(1-\a_\fm/2)},\quad\eta_\fm=\frac{\a_\sf}{\a_\fm}\eta_\sf
\eea
where $\a_\fm$ is given implicitly as the solution of a cubic equation.

\subsection{Derivation of the formulae for  $\L_\sn,\L_\sh,\L_\fm$}\label{sec:deriv}

The solution makes use of the identity
\beq\label{eq:ux}
\lim_{t\to\infty}\E[(u_t-x_t)^2]=\frac{2(1-a)\s_E^2}{(1-a\a)(1+\a)}+\frac{c^2}{(1-a)^2(1-\a)^2}
\eeq
which holds for $u_t$ and $x_t$ satisfying $u_{t+1}=\a u_t+(1-\a)x_t$ and $x_{t+1}=ax_t+b_t+ct$ with $b_t\sim\N((1-a^2)\s_E^2)$. 

\subsubsection{Maturation}

Let $n_{\ss,t}(\g)=N_{\ss,t}(\g)/N_{\ss,t}$ and $m_{\ss,t}(\g)=M_{\ss,t}(\g)/M_{\ss,t}$ where $N_{\ss,t}=\int d\g N_{\ss,t}(\g)$ and $M_{\ss,t}=\int d\g M_{\ss,t}(\g)$ are the total numbers of immature and mature individuals at generation $t$ for $\ss=\sn,\sh,\sm,\sf$. We make the ans\"atze
\beq
n_{\ss,t}(\g)=G_{\vs_{\ss,t}^2}(\g-u_{\ss,t}),\quad m_{\ss,t}(\g)=G_{\varrho_{\ss,t}^2}(\g-v_{\ss,t}).
\eeq
Given $D_\ss(\phi|\g)=G_{\s_{D,\ss}^2}(\g-x_t)$ and $S(\checkmark|\phi,x_t)=(2\pi\s_S^2)^{1/2} G_{\s_S^2}(\phi-x_t)$,
we have
\beq
m_{\ss,t}(\g)=\frac{1}{W_{\ss,t}}\int d\phi S(\checkmark|\phi,x_t) D_\ss(\g|x_t)n_{\ss,t}(\g)
\eeq
with
\bea
W_{\ss,t}&=&(2\pi\s_S^2)^{1/2}G_{\s_S^2+\s_{D,\ss}^2+\vs_{\ss,t}^2}(u_{\ss,t}-x_t)\\
v_{\ss,t}&=& \a_{\ss,t}u_{\ss,t}+(1-\a_{\ss,t})x_t\\
\varrho_{\ss,t}^2&=&\a_{\ss,t}\vs_{\ss,t}^2\label{eq:varrho}\\
\a_{\ss,t}&=&\frac{\s_S^2+\s_{D,\ss}^2}{\s_S^2+\s_{D,\ss}^2+\vs_{\ss,t}^2}\label{eq:alpha}
\eea

\subsubsection{Asexual reproduction}

$n_{t+1,\sn}(\g')=k^{-1}W_{\sn,t}^{-1}\int d\g H_\sn(\g'|\g)m_{\sn,t}(\g)$ with $H_\sn(\g'|\g)=G_{\s_{H,\sn}^2}(\g'-\g)$ so we have 
\bea
u_{\sn,t+1}&=&v_{\sn,t},\\
\vs^2_{\sn,t+1}&=&\varrho^2_{\sn,t}+\s_{H,\sn}^2
\eea
The genetic variance $\vs_{\sn,t}^2$ reaches a fixed point $\vs_{\sn}^2=\s_{H,\sn}^2/(1-\a_{\sn})$ with $\a_\sn$ given by
\beq
\a_\sn=\frac{\s_S^2+\s_\sn^2}{\s_S^2+\s_\sn^2+\s_{H,\sn}^2/(1-\a_\sn)}.
\eeq
This is a quadratic equation for $\a_\sn$ whose solution is given by Eq.~\eqref{eq:asn}. We have
\bea\label{eq:Lsn}
\L_\sn&=&\ln k +\lim_{t\to\infty}\E[\ln W_{\sn,t}]\\
&=&\ln k +\frac{1}{2}\ln\frac{\s_S^2}{\s_S^2+\s_\sn^2+\vs_\sn^2}-\frac{1}{2}\frac{\lim_{t\to\infty}\E[(u_{\sn,t}-x_t)^2]}{\s_S^2+\s_\sn^2+\vs_\sn^2}.\nonumber
\eea
Using Eq.~\eqref{eq:ux}, this leads to $\L_\sn=\ln k+L(\alpha_\sn,\eta_\sn,a,\s_E^2/\s_S^2,c)$.


\subsubsection{Monoecious sexual reproduction}

$n_{t+1,\sh}(\g')=k^{-1}W_{\sh,t}^{-1}\int d\g H_\sh(\g'|\g_\sf,\g_\sm)m_{\sh,t}(\g_\sf)m_{\sh,t}(\g_\sm)$ with $H_\sh(\g'|\g_\sf,\g_\sm)=G_{\s_H^2}(\g'-(\g_\sf+\g_\sm)/2)$ so we have 
\bea
u_{\sh,t+1}&=&v_{\sh,t},\\
\vs^2_{\sh,t+1}&=&\varrho^2_{\sh,t}/2+\s_{H,\sh}^2
\eea
The genetic variance $\vs_{\sh,t}^2$ reaches a fixed point $\vs_{\sh}^2=\s_{H,\sn}^2/(1-\a_{\sh}/2)$ with $\a_\sh$ given by
\beq
\a_\sh=\frac{\s_S^2+\s_\sh^2}{\s_S^2+\s_\sh^2+\s_{H,\sh}^2/(1-\a_\sh/2)}
\eeq
This is a quadratic equation for $\a_\sh$ whose solution is given by Eq.~\eqref{eq:ash}. We have
\bea\label{eq:Lsh}
\L_\sh&=&\ln k+\lim_{t\to\infty}\E[\ln W_{\sh,t}]\\
&=&\ln k +\frac{1}{2}\ln\frac{\s_S^2}{\s_S^2+\s_\sh^2+\vs_\sh^2}-\frac{1}{2}\frac{\lim_{t\to\infty}\E[(u_{\sh,t}-x_t)^2]}{\s_S^2+\s_\sh^2+\vs_\sh^2}\nonumber
\eea
which using Eq.~\eqref{eq:ux} leads to $\L_\sh=\ln k+L(\alpha_\sh,\eta_\sh,a,\s_E^2/\s_S^2,c)$.

\subsubsection{Sexual reproduction}

$n_{t+1,\fm}(\g')=k^{-1}W_{\sf,t}^{-1}\int d\g H_\sh(\g'|\g_\sf,\g_\sm)m_{\sf,t}(\g_\sf)m_{\sm,t}(\g_\sm)$ with $H_\sh(\g'|\g_\sf,\g_\sm)=G_{\s_H^2}(\g'-(\g_\sf+\g_\sm)/2)$ so we have 
\bea
u_{\fm,t+1}&=&(v_{\sf,t}+v_{\sm,t})/2,\\
\vs^2_{\fm,t+1}&=&(\varrho^2_{\sf,t}+\varrho^2_{\sm,t})/4+\s_{H,\fm}^2
\eea
where $\s_{H,\fm}^2=\s_M^2+\s_R^2$. The genetic variance $\vs_{\fm,t}^2$ reaches a fixed point $\vs_{\fm}^2=\s_{H,\fm}^2/(1-\a_{\fm}/2)$ with $\a_\fm$ given by
\bea
\a_\fm=\frac{1}{2}\left(\a_\sf+\a_\sm\right),\\
\a_\sf=\frac{\s_S^2+\s_{D,\sf}^2}{\s_S^2+\s_{D,\sf}^2+\s_{H,\fm}^2/(1-\a_{\fm}/2)},\\
\a_\sm=\frac{\s_S^2+\s_{D,\sm}^2}{\s_S^2+\s_{D,\sm}^2+\s_{H,\fm}^2/(1-\a_{\fm}/2)}
\eea
This is a cubic equation for $\a_\sm$. We have
\bea\label{eq:Lfm}
\L_\fm&=&\ln \frac{k}{2}+\lim_{t\to\infty}\E[\ln W_{\fm,t}]\\
&=&\ln \frac{k}{2} +\frac{1}{2}\ln\frac{\s_S^2}{\s_S^2+\s_{D,\sf}^2+\vs_\sf^2}-\frac{1}{2}\frac{\lim_{t\to\infty}\E[(u_{\sf,t}-x_t)^2]}{\s_S^2+\s_{D,\sf}^2+\vs_\sf^2}\nonumber
\eea
which using Eq.~\eqref{eq:ux} leads to $\L_\fm=\ln (k/2)+L(\alpha_\fm,\eta_\fm,a,\s_E^2/\s_S^2,c)$.

\begin{figure}[t]
\begin{center}
\includegraphics[width=\linewidth]{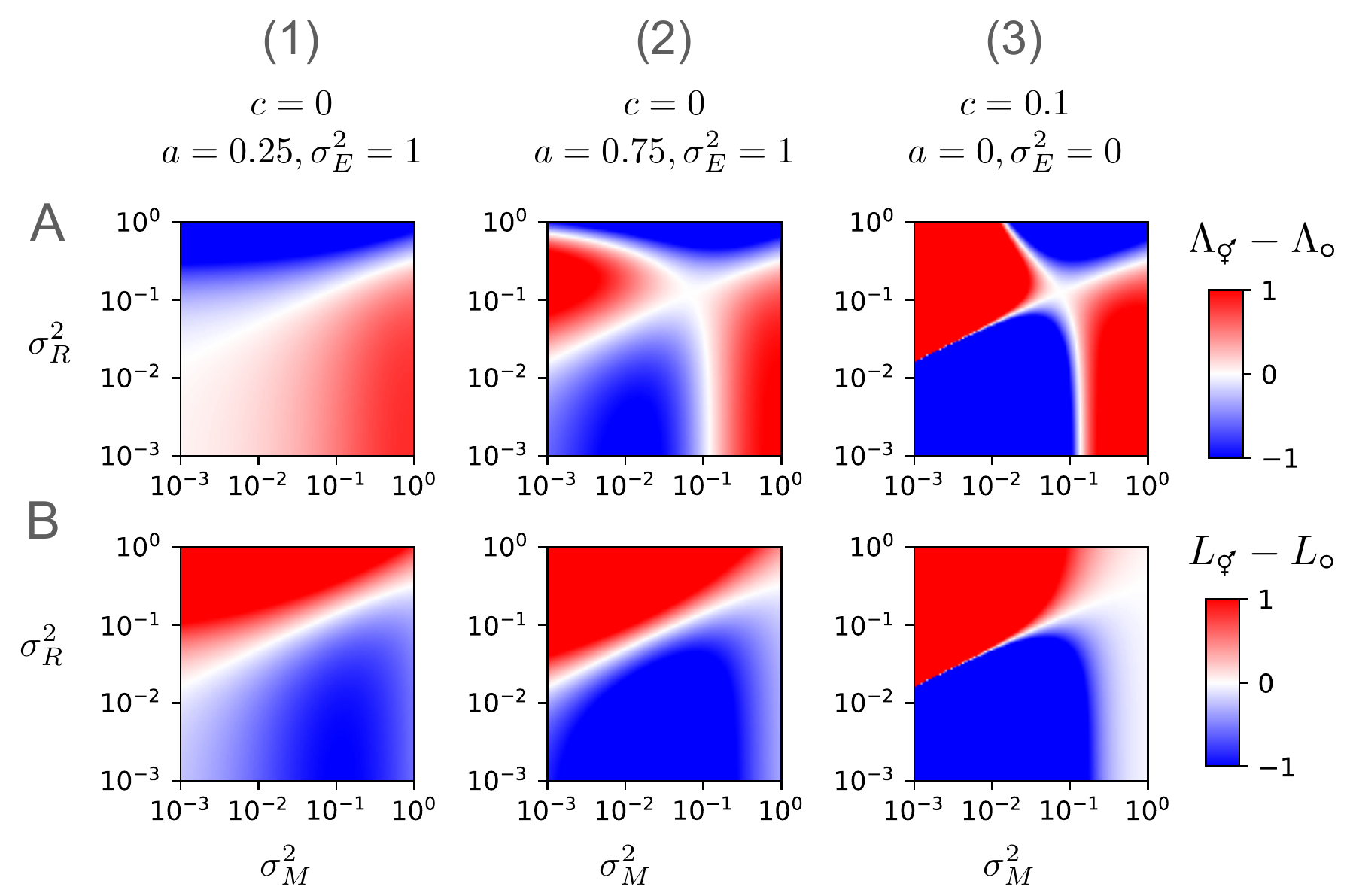}
\caption{{\col {\bf A.} As in Fig.~\ref{fig:AH_opt}A, difference $\L_\sh-\L_\sn$ between the growth rate of sexual and asexual populations as a function of the mutational variance $\s_M^2$ and the segregation variance $\s_R^2$ for three different dynamics of the environment. {\bf B.} Contribution of the mutational load $L_\sh-L_\sn$ defined in Eq.~\eqref{eq:mutload} to the difference $\L_\sh-\L_\sn$, showing in particular that it does not explain the results in condition (1).}\label{fig:mutload}}
\end{center} 
\end{figure}

{\col \subsection{Mutational load}\label{app:mutload}

As seen in Eqs.~\eqref{eq:Lsn}-\eqref{eq:Lsh}-\eqref{eq:Lfm}, the growth rate is generally the sum of three terms,
\beq
\L_\ss=\ln k_\ss +\frac{1}{2}\ln\frac{\s_S^2}{\s_S^2+\s_\ss^2+\vs_\ss^2}+L_\ss
\eeq
where
\beq\label{eq:mutload}
L_\ss=-\frac{1}{2}\frac{\lim_{t\to\infty}\E[(u_{\ss,t}-x_t)^2]}{\s_S^2+\s_\ss^2+\vs_\ss^2}
\eeq
reports the cost due to the lag between the mean trait $u_{\ss,t}$ and the optimal trait $x_t$, which is called the mutational load. We show in Fig.~\ref{fig:mutload} how it contributes to the results of Fig.~\ref{fig:AH_opt}A.
}

\subsection{Joint optimization over mutational, segregational and developmental variances}\label{sec:optima}

Here we show that if we optimize over all the internal parameters that are in principle subject to evolution, namely $\s_M^2$, $\s_R^2$ and $\s_D^2$, then asexual and sexual reproductions lead to identical growth rates. Any difference must therefore rely on constraints on these parameters. 

Taking without loss of generality $\s_S^2=1$ and $k=1$, we have
\bea
&\sup&_{\s_H^2,\s_\sn^2}\L_\sn(\s_H^2,\s_\sn^2,a,\s_E^2,c)\nonumber\\
=&\sup&_{\s_H^2,\s_\sh^2}\L_\sh(\s_H^2,\s_\sh^2,a,\s_E^2)\\
=&\sup&_{(\a,\eta)\in[0,1]^2}L(\a,\eta,a,\s_E^2,c)\nonumber
\eea
since $\L_\ss=L(\a_\ss,\eta_\ss,a,\s_E^2)$ for $\ss=\sn$ and $\ss=\sh$ where in both cases $\eta_\ss$ spans $[0,1]$ when varying $\s_{D,\ss}^2$ in $[0,\infty[$ and, given $\s_{D,\ss}^2$, $\a_\ss$ spans $[0,1]$ when varying $\s_H^2$. Besides, the optimal values $\hat\s_\sn^2$ and $\hat\s_\sh^2$ are identical. The values of $\hat\a_\sn$ and $\hat\a_\sh$ are also identical, which corresponds to values of optimal of $\s_H^2$ that generally differ but are non-zero for the same range of environmental parameters ($\s_H^2=\s_M^2$ in the asexual case and $\s_H^2=\s_M^2+\s_R^2$ in the sexual case).

Similarly,
\bea
&\sup_{\s_H^2,\s_{D,\sf}^2,\s_{D,\sm}^2}\L_\fm(\s_H^2,\s_{D,\sf}^2,\s_{D,\sm}^2,a,\s_E^2,c)\nonumber\\
&=\sup_{(\a,\eta)\in[0,1]^2}L(\a,\eta,a,\s_E^2,c)-\ln 2
\eea
since $\eta_\fm$ spans $[0,1]$ when varying $\s_{D,\sf}^2,\s_{D,\sm}^2$ in $[0,\infty[^2$ at any value of $\s_H^2$ and $\a_\fm$ spans $[0,1]$ when varying $\s_H^2$ in $[0,\infty[$ at any value of $\s_{D,\sf}^2,\s_{D,\sm}^2$. 
The environments in which $\s_H^2,\s_{D,\sf}^2,\s_{D,\sm}^2$ can be non-zero is again identical to the asexual case but now the maximum can be reached for several values of the variables. The main difference with the sexual and monoecious cases is, however, the term $-\ln 2$, which corresponds to the two-fold cost of males.

\subsection{Scaling limit $\s_M^2+\s_R^2\to 0$}

\subsubsection{Scaling limit of the growth rates $\L_\ss$}

Let $\s_H^2=\s_M^2$ in the asexual case and $\s_H^2=\s_M^2+\s_R^2$ in the sexual case. The limit $\s_H^2\to 0$ corresponds to the limit $\a\to 1$. Taking $\epsilon=1-\a$ as small parameter we have $L(\a=1-\epsilon,\eta,a,\s_E^2,c)=L_0(\epsilon,\eta,a,\s_E^2,c)+o(\e)$ with
\bea
&L_0(\epsilon,\eta,a,\s_E^2,c)=\frac{1}{2}\left(\ln\eta-\eta\s_E^2\right)\nonumber\\
&-\frac{1}{2}\left(1-\frac{1}{2}\frac{1+a}{1-a}\eta\s_E^2\right)\epsilon-\frac{1}{2}\eta\frac{c^2}{(1-a)^2}(1-\e)\e^{-2}
\eea
This quantity diverges when $\e\to 0$ if $c>0$ which corresponds to the fact that a finite mutational variance is necessary to cope with a systematically changing environments. Only if $c$ scales with $\e^r$ and $r\geq 2$, is it possible to sustain such a change with a vanishing mutational variance, a situation that may arise if $c$ is a mutational bias that itself vanishes with the rate of mutations.\\

$\e$ has different scalings with $\s_M^2$ and $\s_R^2$ depending on the mode of reproduction:
\bea
\e_\sn&\sim& \eta_\sn^{1/2}\s_M\\
\e_\sh&\sim&2\eta_\sh(\s_M^2+\s_R^2)\\
\e_\fm&\sim&(\eta_\sf+\eta_\sm)(\s_M^2+\s_R^2)\
\eea

\subsubsection{Scaling limit of the difference $\L_\sh-\L_\sn$}\label{sec:sh0}

In this limit, $\s_G^2\sim \eta_\sh^{1/2}\s_M/2$ so $\L_\sh>\L_\sn$ if $\s_R^2\sim \s_M^q$ with $q<2$ and $\L_\sh<\L_\sn$ if $\s_R^2\sim \s_M^q$ with $q>2$. For $c=0$, the sign of $1-(1/2)(1+a)/(1-a)\eta\s_E^2$ also matters, which corresponds to the condition for $\hat\s_M^2=0$ and the qualitative difference between conditions (1) and (2) in Fig.~\ref{fig:AH_opt}.

\subsubsection{Scaling limit of the optimal developmental variances $\s_{D,\sf}^2$ and $\s_{D,\sm}^2$ when $c=0$}

For $\ss=\fm$ and $c=0$, we have
\bea
&\L_\fm\simeq\frac{1}{2}(\ln\eta_\sf-\eta_\sf\s_E^2)-\ln 2-\frac{1}{2}\large[(1-\eta_\sf\s_E^2)(\eta_\sf-\eta_\sm)\nonumber\\
&+(1-\frac{1}{2}\frac{1+a}{1-a}\eta_\sf\s_E^2)(\eta_\sf+\eta_\sm)\large]\s_H^2
\eea
To leading order in $\s_H^2$, $\L_\fm$ depends on $\eta_\sf$ but not on $\eta_\sh$ with the optimal value $\hat\s_{D,\sf}^2=\max (0,\s_E^2-1)$. To first order in $\s_H^2$, $\partial\L_\fm/\partial\eta_\sm(\hat\eta_\sf)=-(1-3a)\min(1,\s_E^2)\s_H^2/[4(1-a)]$ so $\hat\s_{D,\sm}^2=0$ if $a>1/3$ and $\hat\s_{D,\sm}^2=\infty$ if $a<1/3$. Effectively, what is needed in this second case is $\eta_\sm\s_H^2\ll 1$ or $\s_H^2\ll 1+\s_{D,\sm}^2$ which does not necessarily requires $\s_{D,\sm}^2$ to be very large when $\s_H^2$ is small. Finally, we note that $\hat\L_\fm-\hat\L_\sn=\hat\L_\fm-\hat\L_\sh=-\ln 2$ when $\s_H^2= 0$.

\subsubsection{Scaling limit $\s_M^2+\s_R^2\to 0$ when $a=0$, $\s_E^2=0$ but $c\neq 0$}\label{sec:csh2scaling}

When $\s_H^2=\s_M^2+\s_R^2$ is small relative to $\s_S^2=1$, and $a=0$, $\s_E^2=0$, we have
\bea
\L_\sh&\simeq &\frac{1}{2}\ln\eta_\sh-\frac{c^2}{8\eta_\sh\s_H^4},\\
\L_\fm&\simeq &\frac{1}{2}\ln\eta_\sf-\frac{\eta_\sf c^2}{2(\eta_\sf+\eta_\sm)^2\s_H^4}-\ln 2
\eea
Since $\eta_\ss=1/(1+\s_{D,\ss}^2)\in ]0,1]$, the maximum of $\L_\sh$ is achieved for $\hat\s_{D,\sh}^2=0$ and the maximum of $\L_\fm$ relative to $\s_{D,\sm}^2$ for $\hat\s_{D,\sm}^2=0$. The maximum of $\L_\fm$ relative to $\s_{D,\sf}^2$ is, on the other hand, non trivial when $c$ is sufficiently large (Appendix~\ref{app:S9}), and scales with $c$ as $\hat\s_{D,\sf}^2\sim c^2/\s_H^4$. More generally, all quantities depend on $c$ and $\s_H^2$ via the $c/\s_H^2$ with for instance the fraction of surviving males scaling as $M_\sh/N_\sh\sim e^{-c^2/\s_H^4}$.

\subsubsection{Scaling limit of $\hat\s_M^2$ when $c\to 0$}\label{sec:cscaling}

The value of $\s_M^2$ that optimize $\L_\sn$ is never zero when $c>0$ but it decreases sharply as $c\to 0$. When $a=0$ and $\s_E^2=0$, we have indeed $\L_\sn=(1/2)\ln(\eta_\sn\alpha_\sn)-\eta_\sn\alpha_\sn c^2/(2(1-\a)^2)$ with $\a_\sn\sim 1-\eta_\sn^{1/2}\s_M$ when $\s_M^2\ll\s_S^2$, so that
\beq
\L_\sn\simeq\frac{1}{2}\ln\eta_\sn-\frac{1}{2}\eta_\sn^{1/2}\s_M-\frac{c^2}{2\s_M^2}
\eeq
The optimum is for $\hat\s_M^2=(4c^2/\eta_\sn)^{2/3}$ showing that $\hat\s_M^2$ decreases with $c$ as $c^{4/3}$.

\section{Threshold values of the segregation variance}

\subsection{Formula for $\s_G^2$}\label{sec:sG2}

A sufficient (but non-necessary) condition for having $\L_\sh=\L_\sn$ when $\s_{D,\sn}^2=\s_{D,\sh}^2$ is that $\a_\sn=\a_\sh$. Given $\b_\sh=(\s_M^2+\s_R^2)/(\s_S^2+\s_D^2)=\b_\sn+\s_R^2/(\s_S^2+\s_D^2)$, this equation can be solved in $\s_R^2$ to obtain
\beq
\s_G^2=\frac{\s_M^2}{4}\left[\sqrt{1+4\frac{\s_S^2+\s_D^2}{\s_M^2}}-1\right].
\eeq
When $\s_R^2=\s_G^2$, it follows from Eq.~\eqref{eq:alpha} that the asexual and sexual populations have identical genetic variances: $\varsigma_\sn^2=\varsigma_\sh^2$.

\subsection{Gaussian allelic approximation}\label{sec:GAA}

One way to achieve $\s_R^2=\s_G^2$ is to assume an infinitesimal model where  $\g=\sum_{i=1}^L\g^i$ where $L$ is the number of loci and $\g^i$ the contribution of the allele at locus $i$. Starting from two parents with alleles $\g_\sf^i$ and $\g_\sm^i$, the process of recombination is assumed to lead to offspring with alleles $\g_o^i$ such that $\g_o^i=\g_\sf^i$ or $\g_o^i=\g_\sm^i$ independently for each $i$ with probability $1/2$ (Mendelian sampling).

This prescription is sufficient to conclude that $\E[\g_o]=(\g_\sf+\g_\sm)/2$ where the expectation is relative to Mendelian sampling conditionally to the values of $\g_\sf$ and $\g_\sm$. For each locus $i$, we have indeed $\E[\g_o^i]=(1/2)\g_\sf^i+(1/2)\g_\sm^i$ and therefore $\E[\g_o]=\E[\sum_i\g_o^i]=\sum_i\E[\g_o^i]=(\g_\sf+\g_\sm)/2$.

It is not sufficient, however, to derive the segregation variance $\s_R^2={\rm Var}[\g_o]=\E[\g_o^2]-\E[\g_o]^2$. We have indeed
\bea
\E[(\g_o^i-\E[\g_o^i])^2]&=&\frac{1}{2}\left(\g_\sf^i-\frac{\g_\sf^i+\g_\sm^i}{2}\right)^2+\frac{1}{2}\left(\g_\sm^i-\frac{\g_\sf^i+\g_\sm^i}{2}\right)^2\nonumber\\
&=&\frac{(\g_\sf^i-\g_\sm^i)^2}{4}
\eea
and therefore
\bea
\E[(\g_o-\E[\g_o])^2]&=&\E\left[\left(\sum_i(\g_o^i-\E[\g_o^i])\right)^2\right]\nonumber\\
&=&\sum_i\E[(\g_o^i-\E[\g_o^i])^2]\\
&=&\sum_i\frac{(\g_\sf^i-\g_\sm^i)^2}{4}\nonumber
\eea
where we use the assumption that alleles are sampled independently.

Here we need the variance of the distribution of alleles, and not just its mean, to conclude. Let $\s_{\ell,\sf}^2$ and $\s_{\ell,\sm}^2$ be these variances, i.e., $\s^2_{\ell,\ss}=\frac{1}{L}\sum_i(\g_\ss^i-\g_\ss)^2$ for $\ss=\sf,\sm$. Then
\bea
\E[(\g_o-\E[\g_o])^2]&=&\frac{1}{4}\sum_i(\g_\sf^i-\g_\sm^i)^2\\
&=&\frac{L}{4}\left((\g_\sf-\g_\sm)^2+\s_{\ell,\sf}^2+\s_{\ell,\sm}^2\right).\nonumber
\eea

We can proceed by making the additional assumption that alleles are themselves distributed normally independently of each other (the Gaussian allelic approximation). If the distribution of parental genotypes in the population of mature individual is itself Gaussian with variances $\varrho_{\sf,t}^2$ and $\varrho_{\sm,t}^2$ (which are identical in the monoecious case), as in our model, then the central limit theorem constrains the variance of the alleles $\s_{\ell,\sf}^2$ and $\s_{\ell,\sm}^2$ to be respectively $\s_{\ell,\sf}^2=L\varrho_{\sf,t}^2$ and $\s_{\ell,\sm}^2=L\varrho_{\sm,t}^2$. In the limit $L\to\infty$, we then have the simple result
\beq
\s_R^2=\frac{\varrho_{\sf,t}^2+\varrho_{\sm,t}^2}{4}.
\eeq
The variances $\varrho_{\ss,t}$ are given in Eq.~\eqref{eq:varrho} by
\beq
\varrho_{\ss,t}^2=\a_{\ss,t}\vs_{\ss,t}^2=\frac{\s_S^2+\s_{D,\ss}^2}{\s_S^2+\s_{D,\ss}^2+\vs_{\ss,t}^2}\vs_{\ss,t}^2
\eeq
In the particular case of monoecious populations, this corresponds to 
\beq\label{eq:sG2}
\s_R^2=\frac{1}{2}\frac{\s_S^2+\s_{D,\sh}^2}{\s_S^2+\s_{D,\sh}^2+\vs_{\sh,t}^2}\vs_{\sh,t}^2
\eeq
Comparing asexually reproducing and monoecious populations with same developmental variance $\s_{D,\sn}^2=\s_{D,\sh}^2$, we have therefore the two recursions
\bea
\vs_{\sn,t+1}^2&=&\frac{\s_S^2+\s_{D}^2}{\s_S^2+\s_{D}^2+\vs_{\ss,t}^2}\vs_{\ss,t}^2+\s_M^2\\
\vs_{\sh,t+1}^2&=&\frac{1}{2}\frac{\s_S^2+\s_{D}^2}{\s_S^2+\s_{D}^2+\vs_{\ss,t}^2}\vs_{\ss,t}^2+\s_M^2+\s_R^2
\eea
which are strictly identical at any generation $t$ when $\s_R^2$ is given by Eq.~\eqref{eq:sG2}. In particular, in $t\to\infty$ limit we obtain again $\vs_{\sn}^2=\vs_{\sh}^2$.

\section{Finite population size effects on male phenotypic variances}\label{sec:finitepop}

For asexual and hermaphroditic population the growth rate $\L_\ss$ with $\ss=\sn,\sh$ can be written as $\L_\ss=\ln k+K_\ss$ where
\bea
&K_{\ss}=\lim_{t\to\infty}\E\left[\ln \frac{M_{\ss,t}}{N_{\ss,t-1}}\right]\\
&=\frac{1}{2}\ln\rho_\ss-\rho_\ss\left[\frac{(1-a)}{(1-a\a_\ss)(1+\a_\ss)}\frac{\s_E^2}{\s_S^2}+\frac{c^2}{2(1-a)^2(1-\a_\ss)^2}\right]\nonumber
\eea
and where $M_{\ss,t}/N_{\ss,t-1}$ represents the fraction of surviving individuals of type $\ss$ at generation $t$,
\beq
\frac{M_{\ss,t}}{N_{\ss,t-1}}=\int d\g d\phi \int d\phi\ S(\checkmark|\phi,x_t)D_\ss(\phi|\g)n_{\ss,{t-1}}(\g).
\eeq

For dioecious populations, the growth rate $\L_\fm$ is controlled by the fraction of surviving females, $\L_\fm=\ln(k/2)+K_\sf$ where
\bea
&K_{\sf}=\lim_{t\to\infty}\E\left[\ln \frac{M_{\sf,t}}{N_{\sf,t-1}}\right]\\
&=\frac{1}{2}\ln\rho_\fm-
\rho_\fm\left[\frac{(1-a)}{(1-a\a_\fm)(1+\a_\fm)}\frac{\s_E^2}{\s_S^2}+\frac{c^2}{2(1-a)^2(1-\a_\fm)^2}\right]\nonumber
\eea
The fraction of surviving males, on the other hand, does not enter explicitly into the growth rate $\L_\fm$ but we can similarly define and compute
\bea
&K_\sm=\lim_{t\to\infty}\E\left[\ln \frac{M_{\sm,t}}{N_{\sm,t-1}}\right]\\
&=\frac{1}{2}\ln\rho_\sm-\rho_\sm\left[\frac{(1-a)}{(1-a\a_\fm)(1+\a_\fm)}\frac{\s_E^2}{\s_S^2}+\frac{c^2}{2(1-a)^2(1-\a_\fm)^2}\right]\nonumber
\eea
with
\beq
\rho_\sm=\frac{\s_S^2}{\s_S^2+\s_{D,\sm}^2+\s_H^2/(1-\a_\fm/2)}.
\eeq

$K_\sm$ controls the typical number of surviving males, which is $e^{K_{\ss}}N$ if $N$ is the typical total number of newly born males. A necessary condition for the population to survive is therefore $K_\sm>-\ln N$ which imposes an upper bound on $\s_{D,\sm}^2$ since $K_\sm\to -\infty$ when $\s_{D,\sm}^2\to 0$.

In the limit $\s_H^2\to 0$ and $c = 0$, we have, with $\s_S^2=1$,
\beq
K_{\ss}\simeq-\frac{1}{2}\left(\ln(1+\s_{D,\ss}^2)+\frac{\s_E^2}{1+\s_{D,\ss}^2}\right)
\eeq
for $\ss=\sn,\sh,\sf,\sm$. Assuming further $\s_{D,\ss}^2\gg\s_E^2$, $K_{\ss}\simeq -\ln\s_{D,\ss}$ and the condition $K_\sm>-\ln N$ becomes $\s_{D,\sm}<N$, or equivalently . Given the assumption $\s_{D,\sm}^2\gg\s_E^2$, this bounds applies whenever $N\gg\s_E$. The phenotypic variance of male is limited by population size with a quadratic scaling: $\s_{D,\sm}^2/\s_S^2<N^2$.

\section{Spatial heterogeneities}\label{sec:spatial}

We consider two types of spatial heterogeneities that we show to be equivalent. First, we allow differences in local environments to contribute to the developmental variance, $\s^2_{D,\rm tot}=\s^2_D+ \s^2_{D, \rm loc}$. Second, we allow different local environments, associated with different optimal phenotypes $y_t$, to enter into the selection $S(\checkmark|\phi,y_t)$, where we assume that these locally optimal phenotypes $y_t$ are distributed normally around a mean optimal value $x_t$ with variance $\s^2_{E,\rm loc}$.

In absence of spatial heterogeneities, we have $M_{t}(\g)=\tilde S(\g,x_t)N_{t}(\g)$ with an effective selection given by
\beq\label{eq:tildeS}
\tilde S(\g,x_t)=(2\pi\s_S^2)^{1/2}G_{\s_S^2+\s_{D}^2}(\g-x_t).
\eeq
In presence of spatial heterogeneities, with $S(\checkmark|\phi,y_t)=(2\pi\s_S^2)^{1/2}G_{\s^2_S}(\phi-y_t)$, $y_t\sim\N(x_t,\s^2_{E,\rm loc})$ and $D(\phi|\g)=G_{\s_{D}^2+\s_{D,\rm loc}^2}(\phi-\g)$, the effective selection becomes
\bea
\tilde S(\g,x_t)&=&\int dy_tG_{\s_{E,\rm loc}^2}(y_t-x_t)\int d\phi S(\checkmark|\phi,y_t)D(\phi|\g)\nonumber\\
&=&(2\pi\s_S^2)^{1/2}G_{\s_S^2+\s_{D}^2+\s_{D,\rm loc}^2+\s_{E, \rm loc}^2}(\g-x_t),
\eea
Introducing $\s_{D,\rm loc}^2$ or $\s_{E, \rm loc}^2$ is therefore formally equivalent to increasing the value of $\s_D^2$. 

\section{Plasticity}

\subsection{Mapping of models with plasticity onto models without plasticity}\label{sec:plast_mapping}

Models where $D(\phi|\g,x_t)=G_{\s^2_{D}}(\phi-(1-\k)\g-\k x_t)$ at a cost $C(\k)$ can formally be mapped onto the basic model by noting that the effective selection on genotypes $\tilde S(\g,x_t)$ in Eq.~\eqref{eq:tildeS} becomes
\bea
\tilde S(\g,x_t)&=&\int d\phi\ C(\k)S(\checkmark|\phi,x_t)D(\phi|\g,x_t)\nonumber\\
&=&(2\pi\s_S^2)^{1/2}\frac{C(\k)}{1-\k}G_{\frac{\s_{D}^2+\s_S^2}{(1-\k)^2}}(\g- x_t).
\eea
The model with phenotypic plasticity is therefore formally equivalent to the basic model with effective parameters
\beq\label{eq:mapping} 
\tilde\s_D^2=\frac{\s_D^2+(1-C(\k)^2)\s_S^2}{(1-\k)^2},\quad \tilde\s_S^2=\frac{C(\k)^2}{(1-\k)^2}\s_S^2.
\eeq

\subsection{Dioecy with plasticity}\label{sec:plasticity}

Generalizing for $c=0$ the derivation of $\L_\fm$ for dioecious reproduction to developmental kernels $D(\phi_\ss|\g_\ss,x_t)=G_{\s_{D,\ss}^2}(\phi_\ss-\l_\ss\g_\ss-\k_\ss x_t)$ for $\ss=\sf,\sm$, we obtain
\bea
&\Lambda_\fm=\ln\frac{k}{2}+\frac{1}{2}\ln(\eta_\fm\a_\fm)\nonumber\\
&-\frac{\l_\sf^2\eta_\fm\a_\fm}{2}\frac{[(\zeta_\fm^2+\zeta_\sf^2)(1+a\a_\fm)-2\zeta_\fm\zeta_\sf(a+\a_\fm)]}{(1-a\a_\fm)(1+\a_\fm)(1-\a_\fm)}\frac{\s_E^2}{\s^2_S}\label{eq:Lplastic}
\eea
with as before
\beq
\a_\fm=\frac{1}{2}(\a_\sf+\a_\sm),\quad\a_\ss=\frac{1}{1+\b_\ss/(1-\a_\fm/2)},\quad\eta_\fm=\frac{\a_\sf}{\a_\fm}\eta_\sf
\eeq
but
\beq
\beta_\ss=\frac{\l^2_\ss\s_H^2}{\s_S^2+\s_{D,\ss}^2},\quad\eta_\ss=\frac{\s_S^2}{\s_S^2+\s_{D,\ss}^2}
\eeq
and
\beq
\zeta_\fm=\frac{1}{2}(\zeta_\sf+\zeta_\sm)+\frac{1}{2}\a_\sm(\zeta_\sf-\zeta_\sm),\qquad \zeta_\ss=\frac{1-\k_\ss}{\l_\ss}
\eeq
for $\ss=\sf,\sm$.\\

When phenotypes are convex combinations of the genotype and the environment, i.e., $\l_\ss+\k_\ss=1$, we have $\zeta_\ss=1$ for $\ss=\sf,\sm,\fm$ and Eq.~\eqref{eq:Lplastic} becomes
\bea
\Lambda_\fm&=&\ln\frac{k}{2}+\frac{1}{2}\ln(\eta_\fm\a_\fm)-\frac{\l_\sf^2\eta_\fm\a_\fm(1-\a_\fm)}{(1-a\a_\fm)(1+\a_\fm)}\frac{\s_E^2}{\s^2_S}\nonumber\\
&=&\ln \frac{k}{2}+L(\alpha_\fm,\eta_\fm,a,\l_\sf^2\s_E^2/\s_S^2)
\eea
which depends on $\l_\sf$ but not on $\l_\sm$. Besides, the results of optimizing with respect to $\s_{D,\sf}^2$ and $\s_{D,\sm}^2$ are obtained from the case without plasticity by rescaling of $\s_E^2$.

More generally, in the limit of small $\s_H^2\to 0$ where $\alpha=1-\epsilon$, we have to first order in $\epsilon$
\beq
(\zeta_\fm^2+\zeta_\sf^2)(1+a\a_\fm)-2\zeta_\fm\zeta_\sf(a+\a_\fm)\simeq 2(1-a)\zeta_\sf^2 \epsilon
\eeq
which is independent on $\l_\sm,\k_\sm$ even if considering $\l_\sf+\k_\sf\neq 1$ and $\l_\sm+\k_\sm\neq 1$.

\subsection{Phenotypic non-equivalence}\label{sec:phenononeq}

The mapping of Eq.~\eqref{eq:mapping} conceals an important difference at the phenotypic level where we have, prior to selection, given $N_{t}(\g)\propto G_{\vs_{t}^2}(\g-u_{t})$,
\beq
\Phi_t(\phi)=\int d\g D(\phi|\g)N_t(\g)\propto G_{\s_{D}^2+(1-\k)^2\vs_t^2}(\phi-\k x_t-(1-\k)u_t)
\eeq
and, after selection,
\bea
&\Phi'_t(\phi)\propto S(\checkmark|\phi,x_t)\Phi_t(\phi)\\
&\propto G_{(\s_S^{-2}+\s_{D,t}^{-2})^{-1}}\left(\phi-\frac{\k\s_S^2+\s_{D,t}^2}{\s_S^2+\s_{D,t}^2} x_t-(1-\k)\frac{\s_S^2}{\s_S^2+\s_{D,t}^2}u_t\right)\nonumber
\eea
where $\s_{D,t}^2=\s_{D}^2+(1-\k)^2\vs_t^2$. So even though the genetic variances may be identical, the phenotypic variances $\s_\Phi^2=(\s_S^{-2}+\s_{D,t}^{-2})^{-1}$ differ depending on the presence or absence of plasticity. For pure plasticity ($\k=1$, $\s_{D}^2=0$) we have $\s_\Phi^2=(\s_S^{-2}+\vs_t^{-2})^{-1}$  while for pure noise ($\k=0$, $\s_{D}^2>0$) we have $\s_\Phi^2=(\s_S^{-2}+\s_{D}^{-2})^{-1}$. This is important for empirical interpretation. Although the increase of pure plasticity is formally equivalent to the increase of pure noise, the \emph{more plastic} sex has the \emph{narrower} phenotypic distribution, while the \emph{more noisy} sex has the \emph{broader} spread of the observed trait.

\section{Numerical simulations}\label{sec:simu}

\subsection{Principles of the simulations for the basic model}

The analytical formulae for the growth rates can be compared to the results of numerical simulations with populations of finite size $N$ over a finite number $T$ of generations. In these simulations, the population $\mathcal{P}_{\ss,t}$ of newly born individuals of type $\ss$ at generation $t$ is described by a list of $N$ genotypes $[\g_1,\dots,\g_N]$, which are arbitrarily taken to be $\g_i=0$ in the initial population ($t=0$). Given $x_{t-1}$, the simulation consists in the iteration of four steps:

1. Environmental update: $x_t=ax_{t-1}+b+ct$ with $b\sim\N((1-a^2)\s_E^2)$

2. Selection: $\mathcal{P}'_{\ss,t},\ W_{\ss,t}=\mathcal{S}[\mathcal{P}_{\ss,t}]$

3. Reproduction: $\mathcal{O}_t,\ W_t=\mathcal{R}[\mathcal{P}'_t]$

4. Normalization: the $N$ elements of $\mathcal{P}_{\ss,t+1}$ are drawn at random with replacement  from $\mathcal{O}_t$

The selection step 2 is similar in all cases: for each $\g\in \mathcal{P}_{\ss,t}$, a phenotype is computed as $\phi=\g+\nu$ where $\nu\sim\N(\s_{D,\ss}^2)$ and $\g$ is included in the list of surviving individuals $\mathcal{P}'_{\ss,t}$ with probability $S(\checkmark|\phi,x_t)$. $W_{\ss,t}$ reports the fraction of surviving individuals.

The reproduction step 3 depends on whether the population reproduces asexually or sexually and, in the second case, whether it is monoecious or dioecious.

For asexual populations, the population $\mathcal{P}'_t$ consists of $\mathcal{P}'_{\sn,t}$ obtained in step 2 and $W_t=W_{\sn,t}$. Each $\g_\sn\in\mathcal{P}'_{\sn,t}$ produces $k=2$ offsprings in $\mathcal{O}_t$ with genotype $\g'=\g_\sn+\nu$ where $\nu\sim\N(\s_M^2)$.

For monoecious populations, the population $\mathcal{P}'_t$ consists of $\mathcal{P}'_{\sh,t_t}$ and $W_t=W_{\sh,t}$. Each $\g_\sf\in\mathcal{P}'_{\sh,t}$ produces $k=2$ offsprings in $\mathcal{O}_t$ with genotype $\g'=(\g_\sf+\g_\sm)/2+\nu$ where $\g_\sm$ is chosen at random in $\mathcal{P}'_{\sh,t}$ and where $\nu\sim\N(\s_M^2+\s_R^2)$.

For dioecious populations, $\mathcal{P}'_t$ consists of both $\mathcal{P}'_{\sf,t}$ and $\mathcal{P}'_{\sm,t}$ and $W_t=W_{\sf,t}$. Each $\g_\sf\in\mathcal{P}'_{\sf,t}$ produces an offspring in $\mathcal{O}_t$ with genotype $\g'=(\g_\sf+\g_\sm)/2+\nu$ where $\g_\sm$ is chosen at random in $\mathcal{P}'_{\sm,t}$ and where $\nu\sim\N(\s_M^2+\s_R^2)$.

Selection may lead to the elimination of all individuals, in which case the simulation is stopped. When this is not the case, the growth rate is estimated as $\Lambda=(\sum_{t=1}^T\ln W_t)/T$ to which a factor $\ln 2$ is subtracted for dioecious populations to take into account the fact that the total population size is $2N$ and not $N$ in this case. The values of $\L$ obtained in this way are consistent with the analytical formulae.

\begin{figure}[t]
\begin{center}
\includegraphics[width=.6\linewidth]{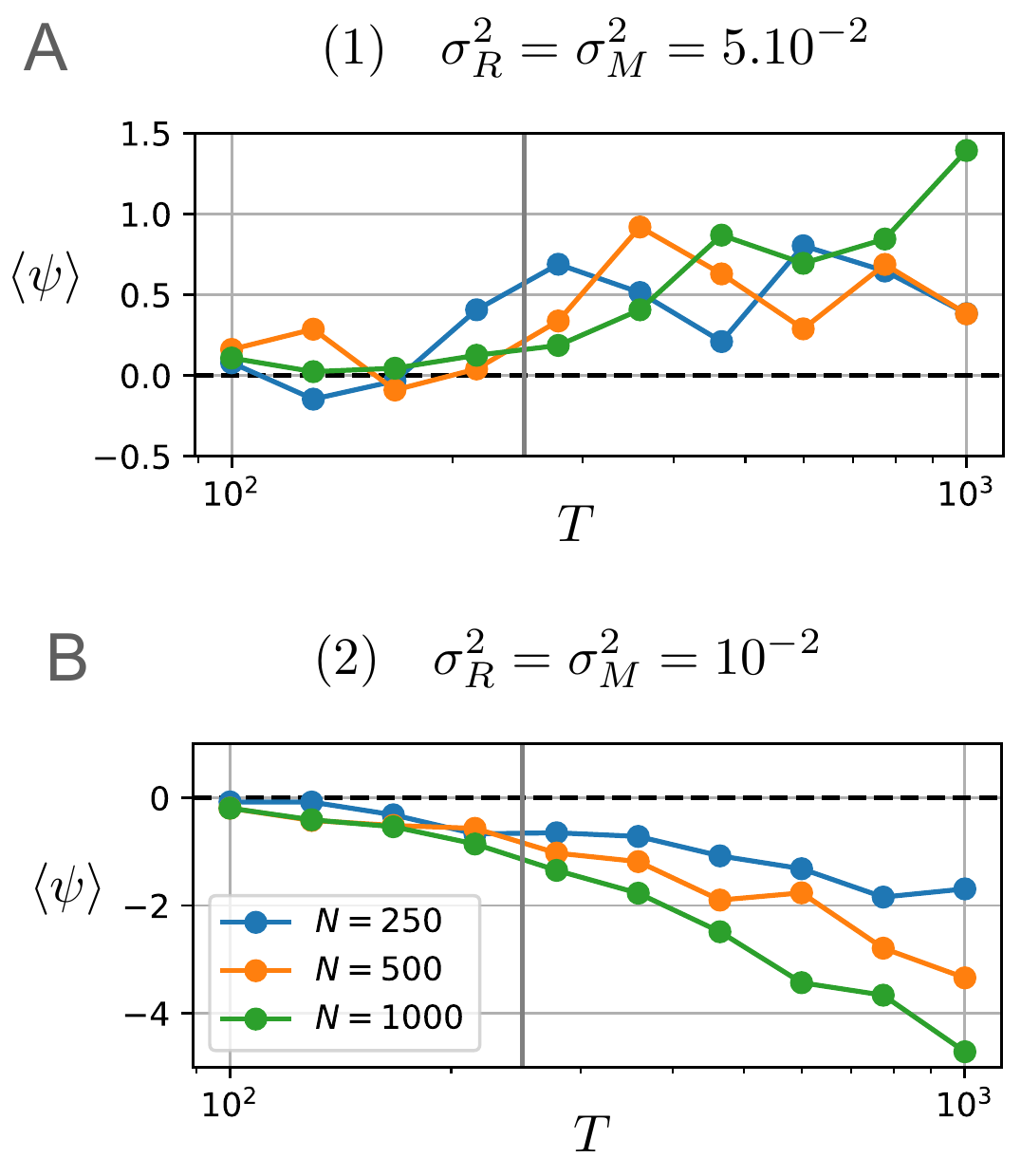}
\caption{Fig.~\ref{fig:AH_opt}C is the result of numerical simulations where the population size is $N=250$ and the number of generations is $T=250$. When increasing these numbers, the mean value of modifier, $\langle\psi\rangle$, takes larger absolute values. {\bf A.} Condition (1) of Fig.~\ref{fig:AH_opt}C at $\s_R^2=\s_M^2=5.10^{-2}$. {\bf B.} Condition (2) of Fig.~\ref{fig:AH_opt}C for $\s_R^2=\s_M^2=10^{-2}$. These values correspond to white zones in Fig.~\ref{fig:AH_opt}C where no selection is apparent. Here, we see that considering a larger number of generations makes $\langle\psi\rangle$ larger in the first case and smaller in the second case, consistent with predictions based on $\L_\sh-\L_\sn$. As in Fig.~\ref{fig:AH_opt}C, these results are averages over 100 simulations. Unsurprisingly, they are more stochastic for smaller population size.\label{fig:evo_TM}}
\end{center} 
\end{figure}

\begin{figure}[t]
\begin{center}
\includegraphics[width=\linewidth]{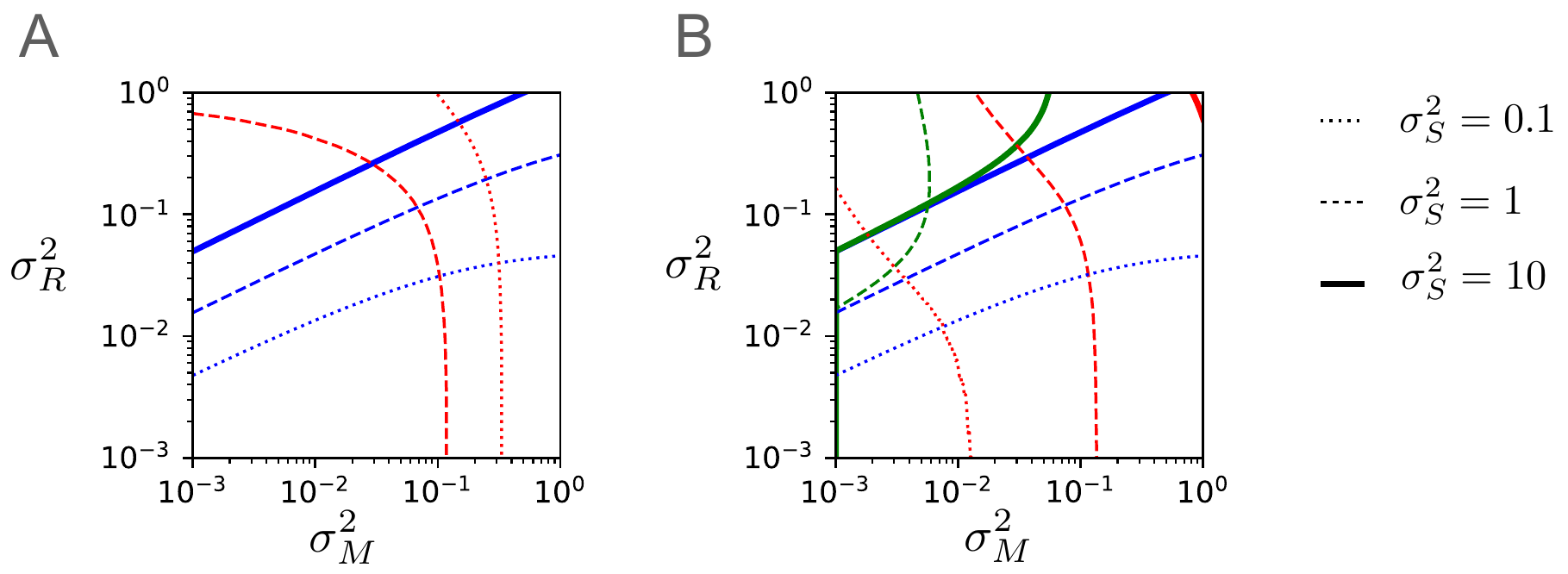}
\caption{{\col {\bf A.} Extension of Fig.~\ref{fig:sM2}B to different values of the stringency of selection $\s_S^2$ (for $\s_D^2=0$). The blue curve corresponds to $\s_G^2$, the value of the segregation variance $\s_R^2$ above which sexual populations have larger genetic variance than asexual populations. The red curve corresponds to $\s_C^2$, such that sex is advantageous when $\s_G^2<\s_R^2<\s_C^2$ or $\s_C^2<\s_R^2<\s_G^2$. The different styles of line correspond to different values of $\s_S^2$: full line for $\s_S^2=10$, dashed line for $\s_S^2=1$ and dotted line for $\s_S^2=0.1$.
{\bf B.} Similar to A but for a directionally varying environment with $c=0.1$, $a=0$, $\s_E^2=0$. The curves for $\s_G^2$, which do not depend on the environment, are identical to A. The curves for $\s_C^2$, on the other hand, differ. Additionally, there are now conditions for which a two-fold cost for sex is overcome ($\L_\sh>\L_\sn+\ln 2$), corresponding to values of $\s_M^2$ on the left side of the green curves. \label{fig:s2S}}}
\end{center} 
\end{figure}

\subsection{Competitions between populations}

When competing two populations with different parameters, for instance an asexually and a sexually reproducing population as in Fig.~\ref{fig:AH_opt}B, we perform independently for each population the step 2 and 3 and then draw the $N$ members of the new generation from the joint set of offsprings $\mathcal{O}^{(1)}_t\cup \mathcal{O}^{(2)}_t$. We then report the fraction of individuals from the first population at the end of the simulation.

\subsection{Numerical simulations with modifiers}

With the modifiers $\psi$ or $\d^\sh,\d^\sn$, the genotype of each individual becomes multidimensional but the same principles apply.

\section{Additional results}

\subsection{Extension of Fig.~\ref{fig:AH_opt}C to different population sizes and numbers of generations}\label{app:evo_TM}

Results extending Fig.~\ref{fig:AH_opt}C to different population sizes and numbers of generations of are shown in Fig.~\ref{fig:evo_TM}.

{\col \subsection{Varying the stringency of selection $\s_S^2$}\label{app:s2S}

Results are presented by default for $\s_S^2=1$. Generalizations to $\s_S^2\neq 1$ are obtained by multiplying all variances by $\s_S^2$. We show in Fig.~\ref{fig:s2S} how this changes the results of Fig.~\ref{fig:sM2}.}

\subsection{Conditions to overcome the two-fold cost of dioecy}.\label{app:2fold}

\begin{figure}[t]
\begin{center}
\includegraphics[width=.9\linewidth]{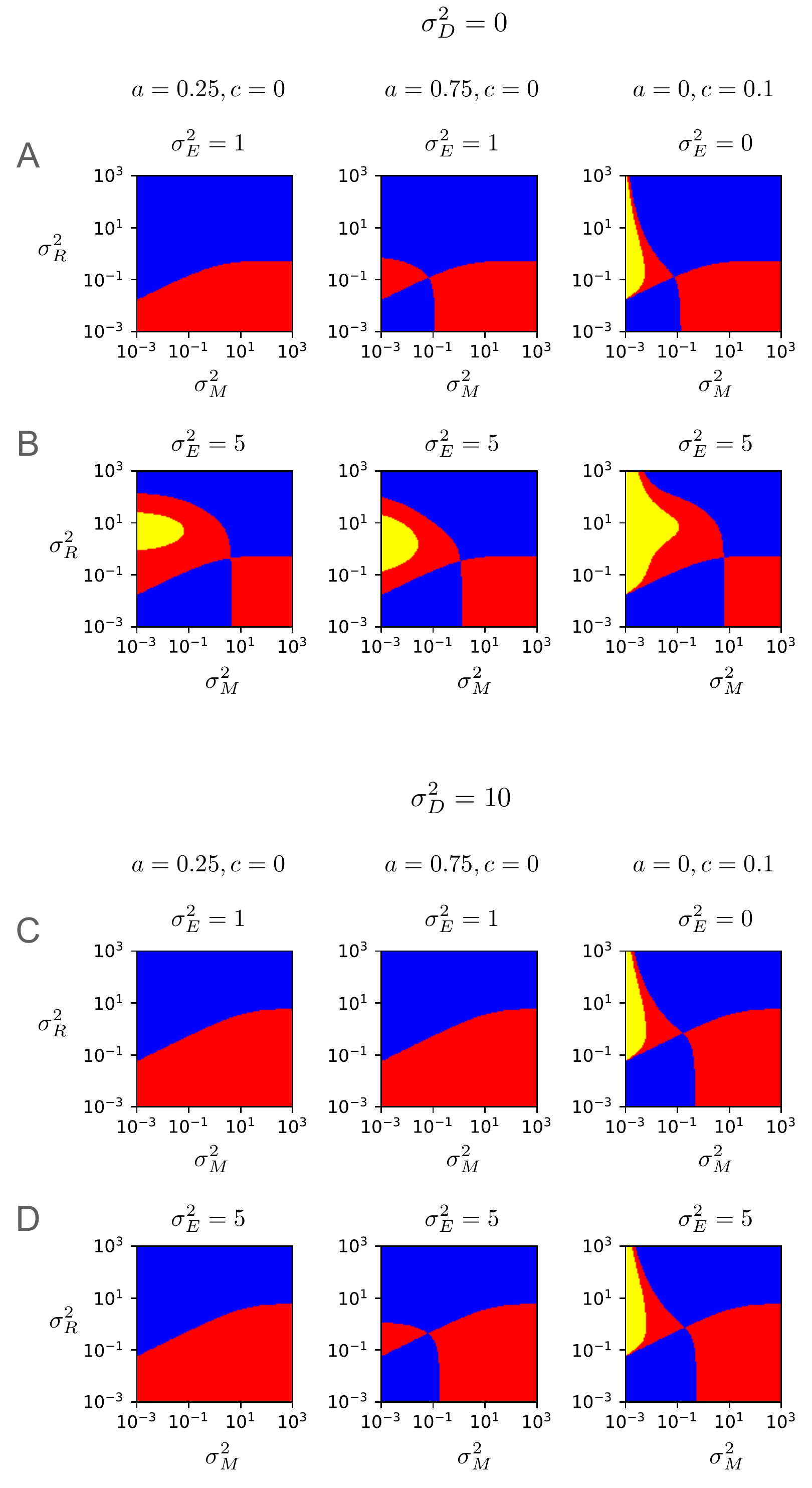}
\caption{For different environmental conditions indicated on the top, values of $\s_M^2,\s_R^2$ for which $\L_\sh<\L_\sn$ in blue, $\L_\fm<\L_\sn<\L_\sh$ in red and $\L_\sn<\L_\fm$ in yellow, given that we always have $\L_\fm=\L_\sh-\ln 2$. The yellow regions thus correspond to genetic and environmental constraints under which the two-fold cost of dioecy is overcome. {\bf A.} Same three conditions as in Fig.~\ref{fig:AH_opt}. {\bf C.} With $\s_D^2=10$ instead of $\s_D^2=0$. {\bf B.} Corresponding conditions with $\s_E^2=5$ and $\s_D^2=0$.  {\bf D.} With $\s_D^2=10$. The two-fold cost of sex is possibly overcome only when $\s_G^2<\s_R^2<\s_C^2$ (see also Fig.~\ref{fig:2fmore}).\label{fig:2fold}}
\end{center} 
\end{figure}

Conditions for which the two-fold of dioecy is overcome are shown in Fig.~\ref{fig:2fold}.

\subsection{Requirements on $\s_R^2$ to overcome the two-fold cost of sex}\label{app:2fmore}

\begin{figure}[t]
\begin{center}
\includegraphics[width=.75\linewidth]{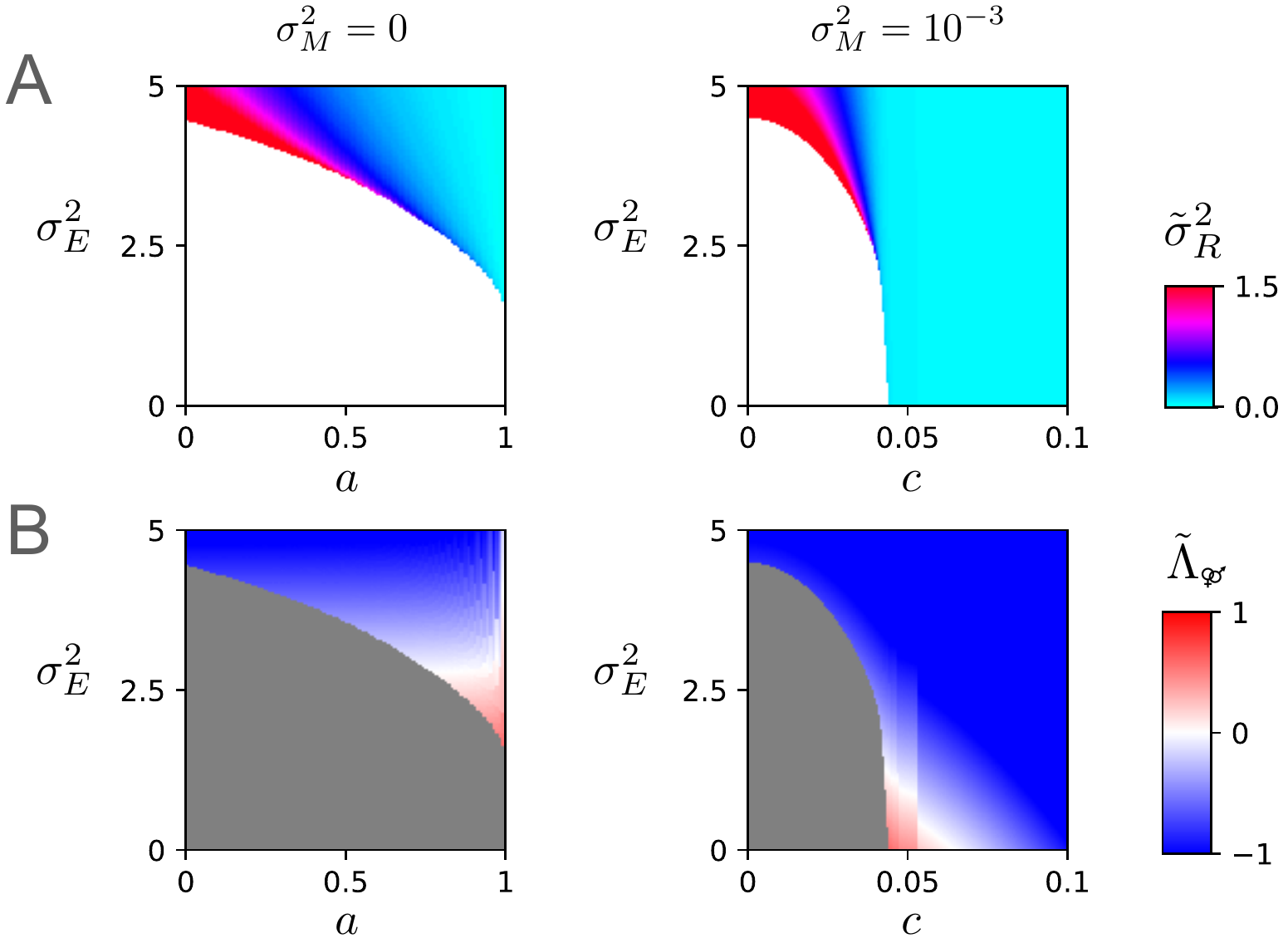}
\caption{{\bf A.} Smallest values of $\s_R^2${\col , denoted $\tilde\s_R^2$,}  at which the two-fold cost of sex is overcome, i.e., $\L_\fm-\L_\sn=\L_\sh-\L_\sn-\ln 2>0$, as a function of $(a,\s_E^2)$ for $c=0$ when $\s_M^2=0$ (left) and as a function of $(c,\s_E^2)$ for $a=0$ when $\s_M^2=10^{-3}$. Environmental conditions for which the two-fold cost is not overcome for any value of $\s_R^2$ are indicated in white. {\bf B.} Values of $\L_\fm$ for the corresponding value of $\s_R^2${\col , denoted $\tilde\L_\fm=\L_\fm(\s_R^2=\tilde\s_R^2)$} (in gray when undefined). Negative {\col $\tilde\L_\fm$}, in blue, correspond to situations where extinction is nearly certain. Here the mean number of offspring is taken to be $k=2$. A larger value of $k$, which corresponds to adding $\ln(k/2)$ to {\col $\tilde\L_\fm$}, would widen the conditions under which survival is possible. \label{fig:2fmore}}
\end{center} 
\end{figure}

The minimal values of $\s_R^2$ at which the two-fold cost of sex is overcome are shown in Fig.~\ref{fig:2fmore}.

\subsection{Evolution of sexual dimorphism under different models for the segregation variance}\label{app:r}

\begin{figure}[t]
\begin{center}
\includegraphics[width=\linewidth]{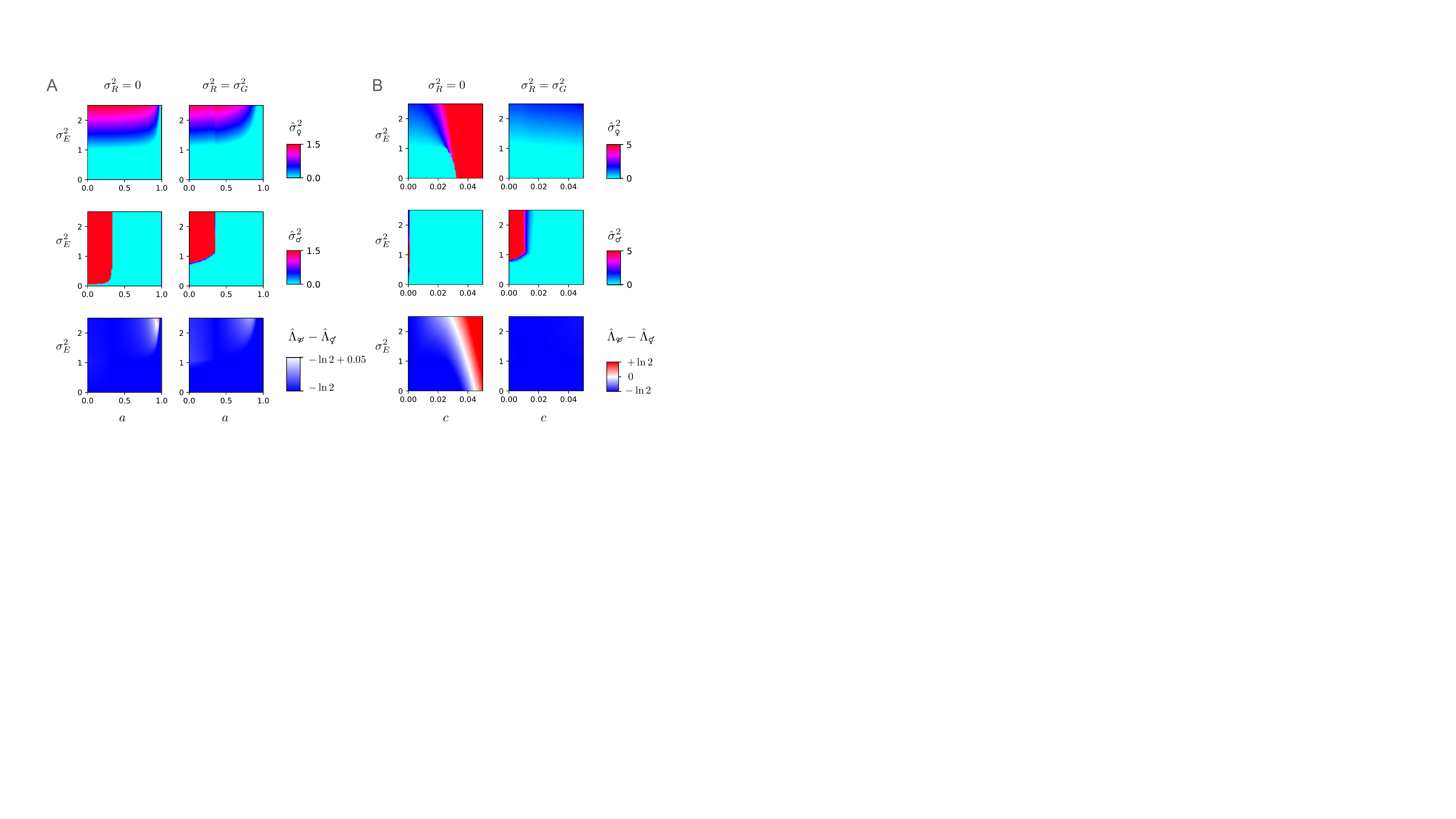}
\caption{{\bf A.} Extension of Fig.~\ref{fig:dimo}A, which corresponds here to the graphs on the left where $\s_M^2+\s_R^2=10^{-2}$, which formally is equivalent to $\s_M^2=10^{-2}$ and $\s_R^2=0$, to the case where $\s_M^2=10^{-2}$ and $\s_R^2=\s_G^2$, where $\s_G^2$ depends on $\s_{\sf}^2$ and $\s_\sm^2$ as indicated in Appendix~\ref{sec:GAA}. The third row reports $\hat \L_\fm-\hat \L_\sh$, the difference of growth rates between dioecious and monoecious populations when optimizing over the developmental variance. This difference is never very far from $-\ln 2$, its value in absence of dimorphism. {\bf B.} Similar to A but as a function of $(c,\s_E^2)$ for $a=0$ instead of as function of $(a,\s_E^2)$ for $c=0$, where $c$ can be interpreted either as a drift of the environment or a mutational bias (Sec.~\ref{sec:bias}). Note the difference of scale compared to A. Most significantly, $\hat \L_\fm-\hat \L_\sh$ can take positive values for sufficiently $c$, indicating that the two-fold cost of males can be overcome through sexual dimorphism.\label{fig:r}}
\end{center} 
\end{figure}

Results extending Fig.~\ref{fig:dimo}A to different values of $\s_M^2$ and $\s_R^2$ are shown in Fig.~\ref{fig:r}.

\subsection{Role of initial conditions in the evolution of sexual dimorphism}\label{app:evomore}

\begin{figure}[t]
\begin{center}
\includegraphics[width=.95\linewidth]{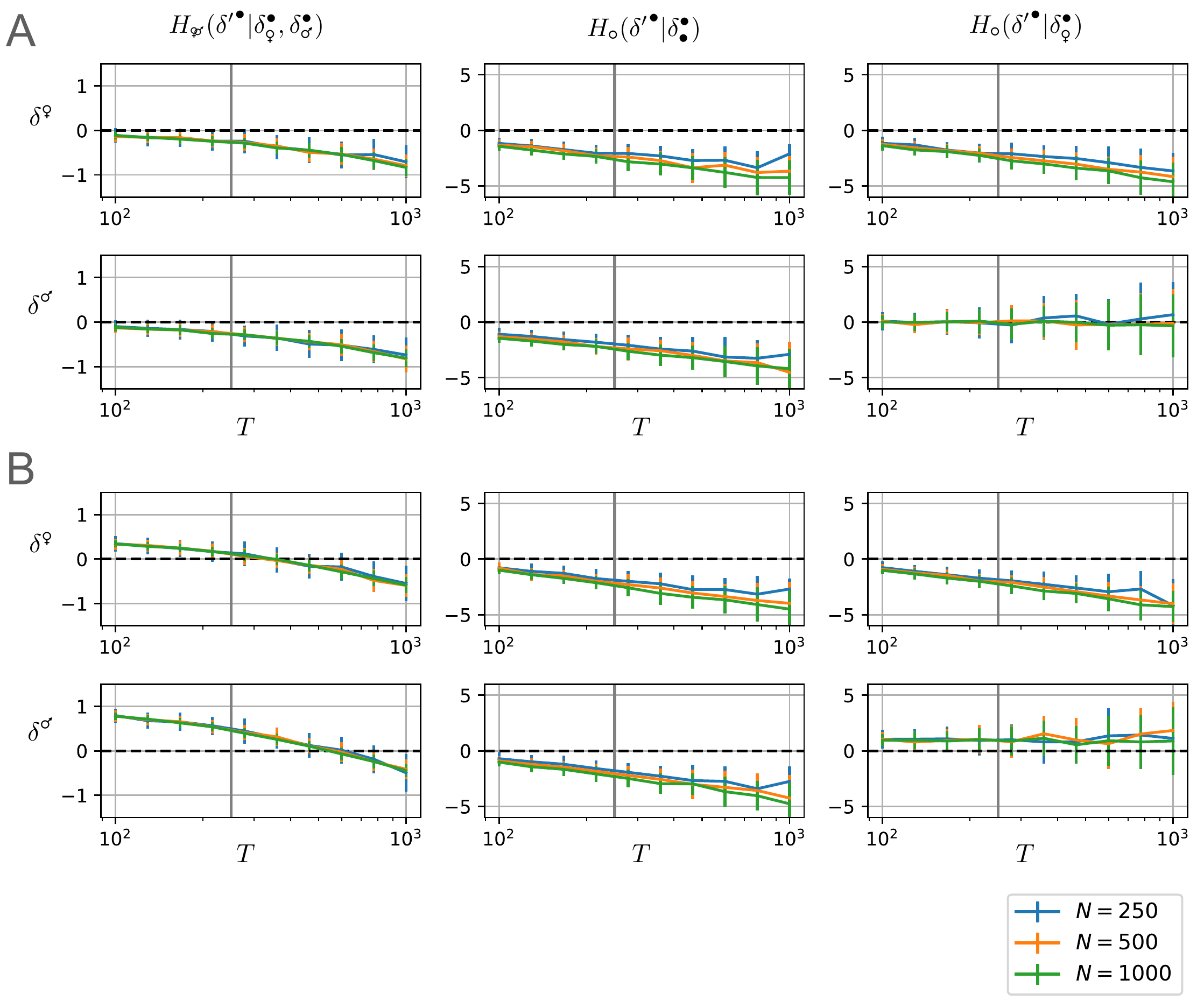}
\caption{Evolution of the modifiers $\d^\sf$ and $\d^\sm$ when considering different modes of transmission and different initial conditions, here illustrated for $a=0.5$ and $\s_E^2=1$. $H({\d'}^\ss|\d_\sf^\ss,\d_\sm^\ss)=H_\fm({\d'}^\ss|\d_\sf^\ss,\d_\sm^\ss)$ assumes that the modifiers $\d^\ss$ are subject to recombination, $H({\d'}^\ss|\d_\sf^\ss,\d_\sm^\ss)=H_\sn({\d'}^\ss|\d_\ss^\ss)$ that they are inherited separately by each sex, and $H({\d'}^\ss|\d_\sf^\ss,\d_\sm^\ss)=H_\sn({\d'}^\ss|\d_\sf^\ss)$ that they are inherited through the females exclusively (as in  Fig.~\ref{fig:dimo}B). {\bf A.} Starting from $\d^\sf=\d^\sm=0$ as in Fig.~\ref{fig:dimo}B, which corresponds to the graphs in the last column. Note the difference of scale on the y-axis in the panels of the first column compared to the others. {\bf B.} Starting from $\d^\sf=\d^\sm=1$, we obtain similar results except for the graph on the bottom right, where, in average over 100 independent simulations, the results essentially reflect the initial conditions. The error bars indicate standard deviations over 100 independent simulations and the different colors correspond to different total population sizes. \label{fig:evomore}}
\end{center} 
\end{figure}

Results extending Fig.~\ref{fig:dimo}B to different initial conditions are shown in Fig.~\ref{fig:evomore}.

\subsection{Evolution of sexual dimorphism under different modes of transmission of the modifiers}\label{app:dimore}

\begin{figure}[t]
\begin{center}
\includegraphics[width=\linewidth]{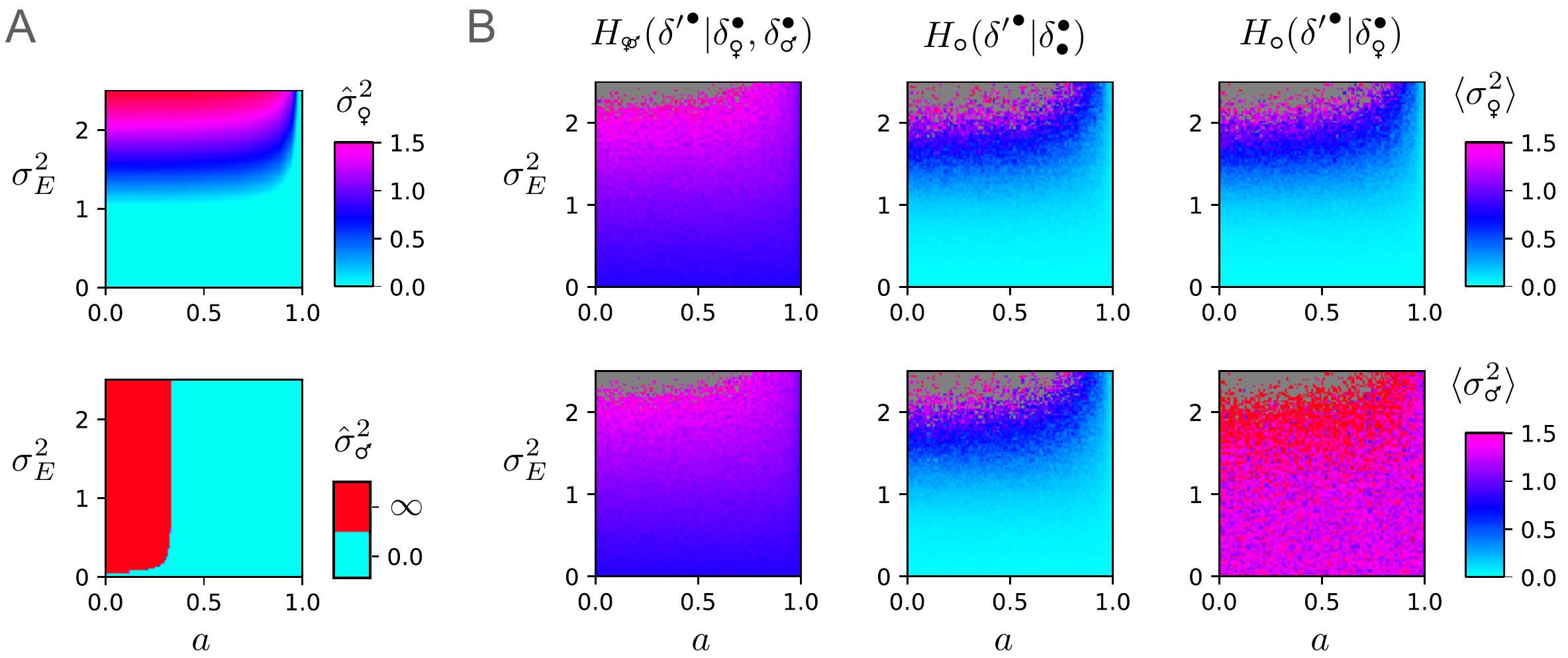}
\caption{Extension of Fig.~\ref{fig:dimo} to different modes of transmission of the modifiers. $H({\d'}^\ss|\d_\sf^\ss,\d_\sm^\ss)=H_\fm({\d'}^\ss|\d_\sf^\ss,\d_\sm^\ss)$ assumes that the modifiers $\d^\ss$ are subject to recombination, $H({\d'}^\ss|\d_\sf^\ss,\d_\sm^\ss)=H_\sn({\d'}^\ss|\d_\ss^\ss)$ that they are inherited separately by each sex, and $H({\d'}^\ss|\d_\sf^\ss,\d_\sm^\ss)=H_\sn({\d'}^\ss|\d_\sf^\ss)$ that they are inherited through the females exclusively (as in  Fig.~\ref{fig:dimo}B). Only in the later case do we observe sexual dimorphism. Note that these results depend on the initial conditions are shown in Fig.~\ref{fig:evomore}.\label{fig:dimore}}
\end{center} 
\end{figure}

Results extending Fig.~\ref{fig:dimo} to different modes of transmission of the modifiers are shown in Fig.~\ref{fig:dimore}.

\subsection{Extension of Fig.~\ref{fig:AH_opt} to directional selection}\label{app:2bis}

\begin{figure}[t]
\begin{center}
\includegraphics[width=.9\linewidth]{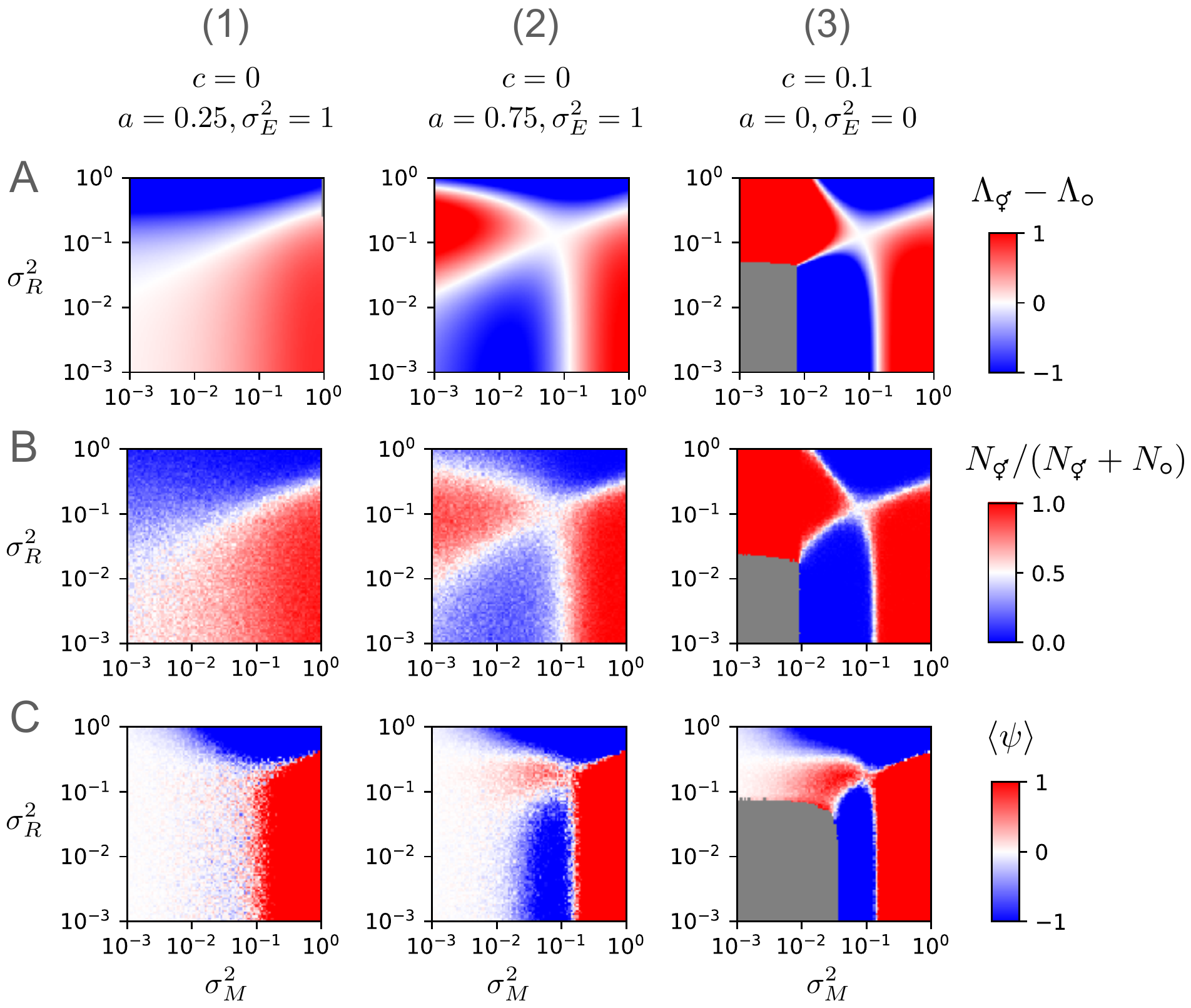}
\caption{Extension of Fig.~\ref{fig:AH_opt} to a third environmental condition where the environment is systematically drifting, $x_{t+1}=c t$ with $c=0.1$. This condition (3) is similar to condition (2). In this case, however, the population may become extinct, which is indicated in gray. In A, the criterion for extinction is $\max(\L_\sh,\L_\sn)<0$. In B and C, it corresponds to cases where more than 10\% of the 100 simulations over which the results are averaged ended up in extinction, i.e., no individual survived after maturation despite a number of newly born individuals maintained to a fixed value, here $N=250$.\label{fig:2bis}}
\end{center} 
\end{figure}

Results extending Fig.~\ref{fig:AH_opt} to an environment that is systematically drifting are shown in Fig.~\ref{fig:2bis}.

\subsection{Optimal mode of reproduction when optimizing over developmental variances}\label{app:2fold2}

\begin{figure}[t]
\begin{center}
\includegraphics[width=\linewidth]{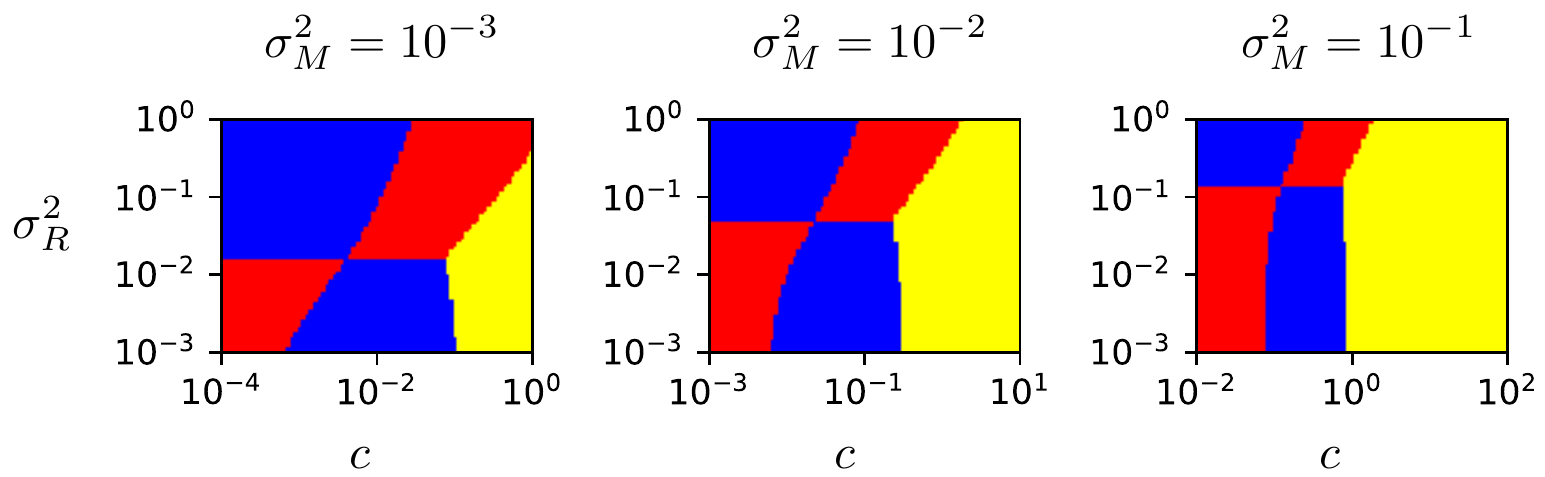}
\caption{Optimal mode of reproduction when optimizing over developmental variances as a function of $c$ and $\s_R^2$ for three values of $\s_M^2$ indicated on the top and $a=0$, $\s_E^2=0$ (note the differences of scales on the x-axes). As in Fig.~\ref{fig:2fmore}, blue indicates that asexual reproduction is optimal, red that it is monoecious sexual reproduction and yellow that it is dioecious sexual reproduction. For large values of $c$, the two-fold cost of males is therefore overcome both relative to asexuality and to monoecy. \label{fig:2fold2}}
\end{center} 
\end{figure}

Results on the optimal mode of reproduction when optimizing over developmental variances are shown in Fig.~\ref{fig:2fold2}.

\subsection{Sexual dimorphism under directional selection}\label{app:S9}

\begin{figure}[t]
\begin{center}
\includegraphics[width=\linewidth]{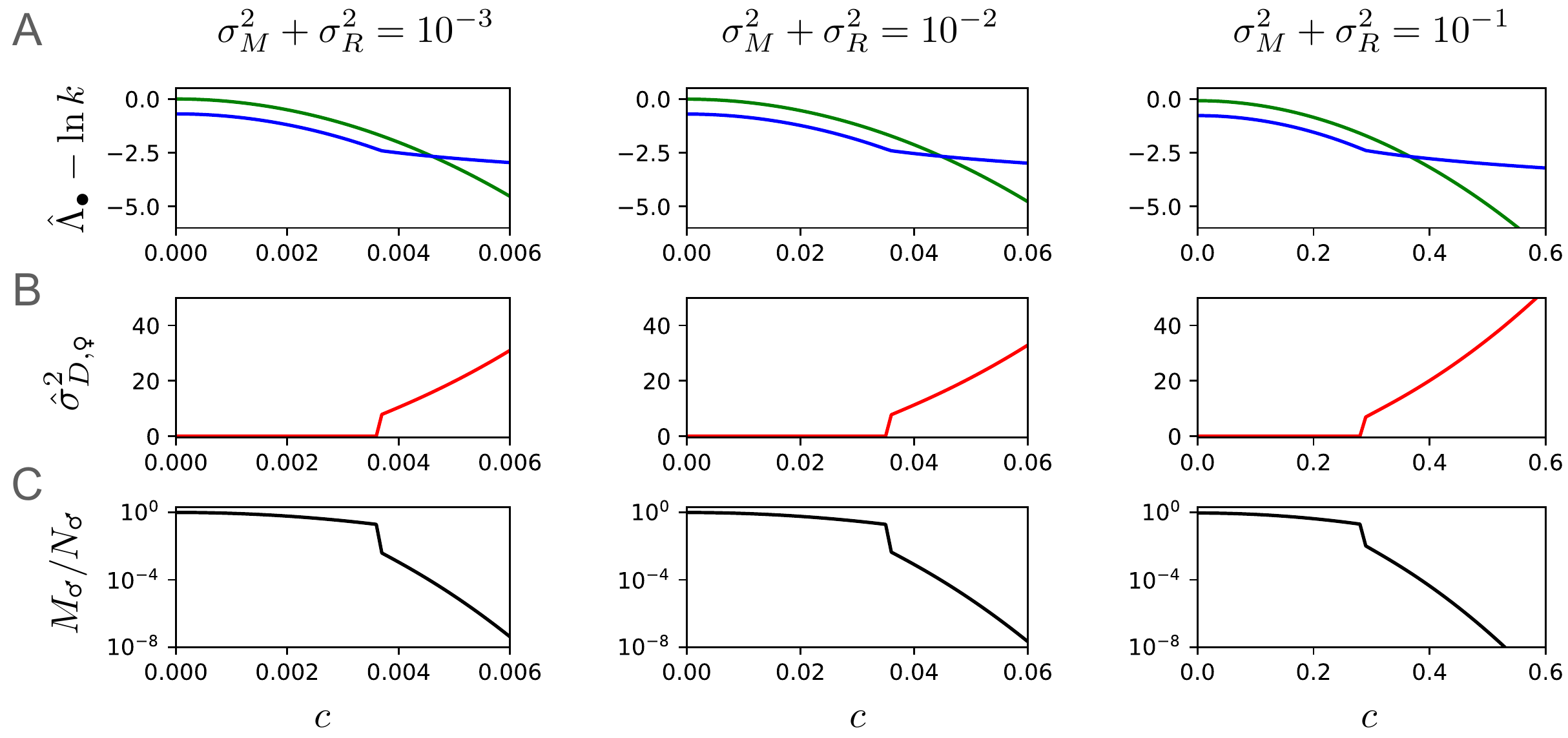}
\caption{Sexual dimorphism for $c>0$, $a=0$, $\s_E^2=0$ and three different values of the mutational and segregation variances indicated on the top. {\bf A.} Optimal growth rates for monoecious (in green) and dioecious (in blue) populations as a function of $c$. The optimization is here performed on the developmental variances. {\bf B.} Optimal female developmental variances $\hat\s_{D,\sf}^2$. In contrast, the optimal developmental variances for monoecious populations and for males in dioecious populations are trivial: $\hat\s_{D,\sh}^2=0$ and $\hat\s_{D,\sm}^2=0$ for any value of $c$ (Appendix~\ref{sec:csh2scaling}). {\bf C.} Mean fraction of males reaching maturation at each generation. Note that this fraction is very small for values of $c$ at which dioecy is advantageous over monoecy (blue curve above the green curve in A). Populations whose size is not significantly larger than the inverse of this ratio may be considered non viable. Finally, note that in the limit, $\s_M^2+\s_R^2\to 0$, the different quantities depend $c$ and $\s_M^2+\s_R^2$ only via $c/(\s_M^2+\s_R^2)$, which explains that the three graphs differ almost only by the scale on their x-axis (Appendix~\ref{sec:csh2scaling}).\label{fig:S9}}
\end{center}
\end{figure}

Results on the optimal degree of sexual dimorphism in directional environments are shown in Fig.~\ref{fig:S9}.

\end{document}